\documentclass[%
 reprint,
superscriptaddress,
 amsmath,amssymb,
 aps,
prl,
]{revtex4-2}

\usepackage{graphicx}% Include figure files
\usepackage{dcolumn}% Align table columns on decimal point
\usepackage{bm}% bold math
\usepackage{braket}
\usepackage{setspace}

\newcommand{\NIST}{
National Institute of Standards and Technology, 325 Broadway, Boulder, Colorado 80305, USA}

\newcommand{\CU}{
Department of Physics, University of Colorado, Boulder, Colorado 80309, USA}
\begin{document}

\preprint{APS/123-QED}
\title{Excited-Band Coherent Delocalization for Improved Optical Lattice Clock Performance}% Force line breaks with 

\author{J.L. Siegel}
\thanks{These authors contributed equally.}
\affiliation{\NIST}
\affiliation{\CU}

\author{W.F. McGrew}%
\thanks{These authors contributed equally.}
\altaffiliation[Present address: ] {IMRA America Inc., Boulder Research Labs, 1551 South Sunset St., Suite C, Longmont, Colorado 80501, USA}%Lines break automatically or can be forced with \\
\affiliation{\NIST}
\affiliation{\CU}

\author{Y.S. Hassan}%
\affiliation{\NIST}

\affiliation{\CU}
\author{C.-C. Chen}%
\affiliation{\NIST}
\affiliation{\CU}
\author{K. Beloy}%
\affiliation{\NIST}
\author{T. Grogan}%
\affiliation{\NIST}

\affiliation{\CU}
\author{X. Zhang}%
\affiliation{\NIST}
\affiliation{\CU}
\altaffiliation[Currently at ]{ ??}%Lines break automatically or can be forced with \\
\author{A.D. Ludlow}%

  \email{andrew.ludlow@nist.gov}

\affiliation{\NIST}

\affiliation{\CU}
\date{\today}% It is always \today, today,
             %  but any date may be explicitly specified

\begin{abstract}
We implement coherent delocalization as a tool for improving the two primary metrics of atomic clock performance: systematic uncertainty and instability.
By decreasing atomic density with coherent delocalization, we suppress cold-collision shifts and two-body losses. 
Atom loss attributed to Landau-Zener tunneling in the ground lattice band would compromise coherent delocalization at low trap depths for our \textsuperscript{171}Yb atoms; hence, we implement for the first time delocalization in excited lattice bands.
Doing so increases the spatial distribution of atoms trapped in the vertically-oriented optical lattice by $\sim7$ times. 
At the same time we observe a reduction of the cold-collision shift by 6.5(8) times, while also making inelastic two-body loss negligible. 
With these advantages, we measure the trap-light-induced quenching rate and natural lifetime of the $^3$P$_0$ excited-state as $5.7(7)\times10^{-4}$ $E_r$\textsuperscript{$-1$}s\textsuperscript{$-1$} and 19(2) s, respectively.

\end{abstract}

%\keywords{Suggested keywords}%Use showkeys class option if keyword
                              %display desired
\maketitle

Optical lattice clocks have emerged on the forefront of frequency metrology, reaching fractional frequency uncertainties in the low-$10^{-18}$ decade \cite{Ushijima2015, McGrew2018, Bothwell2019}.
This high performance has already enabled sensitive explorations of dark matter models \cite{Wciso2018, Wciso2017, Beloy2021, Derevianko2014} and early studies of Earth's geopotential \cite{Chou2010, McGrew2018, Grotti2018, Takamoto2020}.
As optical lattice clocks continue to improve, they promise to surpass classical geodetic measurement \cite{Mehlstaubler2018, Bondarescu2012, Denker2018}, to detect gravitational waves \cite{Kolkowitz2016, Su2018}, and to more deeply probe beyond-Standard-Model physics \cite{Safronova2019, Barontini2021}.

One important systematic effect afflicting lattice clocks is the cold-collision shift.
This density-dependent frequency shift is typically suppressed by exploiting low temperatures and Fermi statistics \cite{Lemke2011}.
Nevertheless, clock transition shifts can still be significant at the $10^{-18}$ level \cite{Lemke2011, Kobayashi2020}, sometimes even when atom number is intentionally restricted to decrease atomic density \cite{Bothwell2019, Ludlow2011}.
Other approaches have been used to reduce the cold-collision shift in specific operational conditions \cite{Akatsuka2010, Ludlow2011, Lemke2011, Aeppli2022, Campbell2017, Okaba2014}, but a simple reduction in atomic density remains a universal and robust strategy to mitigate the effect. 
However, the push for improved clock stability represents a strong competing interest, since higher atom numbers benefit the quantum projection noise (QPN) stability limit for uncorrelated atoms \cite{Itano1993}.
Density-dependent two-body losses also cause excess atom loss \cite{Bishof2011, Ludlow2011}, subsequently degrading stability.

Here, we adapt coherent delocalization, a Floquet engineering method developed for gravimetry \cite{Alberti2009, Ivanov2008}, to reduce the burden of density-dependent effects on optical lattice clocks. 
Amplitude modulation (AM) of a one-dimensional optical lattice at multiples of the Bloch frequency induces tunneling between lattice sites.
It has been shown to increase the root-mean-square spatial extent of trapped \textsuperscript{88}Sr atoms by as much as 15 times in one second of modulation \cite{Ivanov2008}. 
However, while the speed of delocalization increases for the shallowest of lattice depths, Landau-Zener (LZ) tunneling can introduce significant atom loss \cite{Landau1932}. 
To mitigate the loss, we induce tunneling between lattice sites in the excited bands of deeper lattice potentials.
Through the application of adiabatic rapid passage (ARP) on the clock transition motional sidebands \cite{Vitanov2001}, excited lattice bands ($n_z > 0$) can be prepared with high purity. 
The excited bands enlarge AM-induced tunneling rates relative to the ground motional band.  
Using this preparation protocol, we show a nearly order-of-magnitude increase of the spatial extent of our atomic sample after one second of coherent delocalization. 
As an immediate benefit of delocalization, we measure a 6.5(8) times reduction in the cold-collision shift that softens the trade-off between high atom numbers and low systematic frequency shifts.
In addition, two-body loss is rendered negligible, allowing us to unambiguously measure the lattice Raman scattering and natural lifetime limits of the excited-state. 

\begin{figure}
    \centering
    \includegraphics[width=0.5\textwidth,height=120pt]{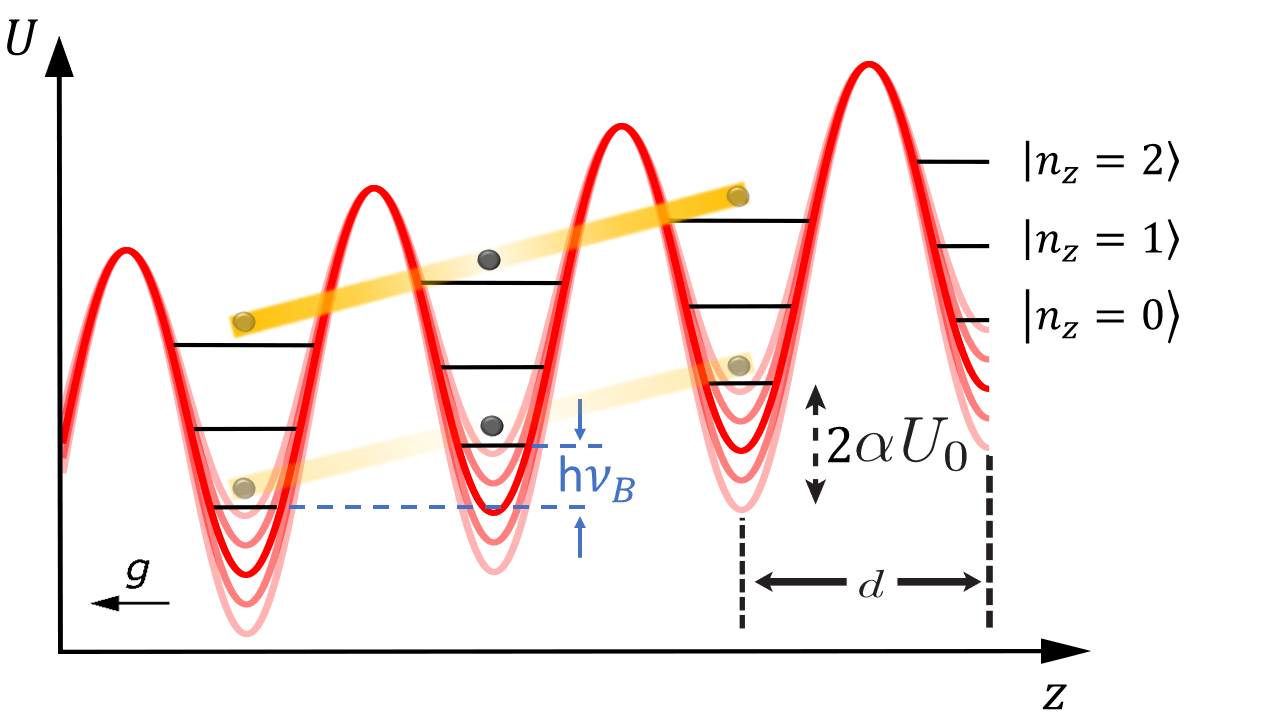}
    \caption{Shaken vertical optical lattice potential. 
    The tunneling rate between neighboring lattice sites, shown as the strength of the yellow bars, increases for higher longitudinal bands. 
  }
    \label{fig:tunnel}
\end{figure}

Vertical lattices use gravity to break the degeneracy between neighboring lattice sites by $h\nu_B$, where $h$ is Plank's constant and $\nu_B \approx 1593$ Hz is the Bloch frequency for \textsuperscript{171}Yb.
As a result, atomic wave functions are localized in Wannier-Stark (WS) states. 
Amplitude modulating the lattice at $\nu_B$ reinstates coherent evolution of the wave function between lattice sites via tunneling, as shown in Fig.~\ref{fig:tunnel}. 
The tunneling rate depends on the overlap of WS wave functions in neighboring lattice sites, which is naturally larger at low trap depths \cite{Alberti2010}. 
We begin by experimentally measuring unwanted atom loss versus lattice depth.  
The main details of our experiment have been described elsewhere \cite{McGrew2018}.
Briefly, our vertically aligned 759-nm magic wavelength optical lattice, with Rayleigh length 2.3 cm, is loaded with up to $10^4$ atoms via two magneto-optical trap (MOT) stages, first using the broad 399-nm transition followed by the narrow 556-nm transition. 
For this work, a Sisyphus cooling mechanism using the clock transition is also applied to reduce the radial atomic temperature to $T_r\sim 450$ nK, the longitudinal atomic temperature to $\sim 600$ nK ($\bar{n}_z = 0.07(3)$), and to enhance loading into an applied lattice depth $U\approx57 ~ E_r$ ($E_r=\frac{\hbar^2k_l^2}{2m}$, where $\hbar=h/2\pi$, $k_l=\frac{2\pi}{\lambda}$, $\lambda$ is the optical lattice wavelength, and $m$ is the mass of the \textsuperscript{171}Yb atom) \cite{Chen2022}. 
We then adiabatically ramp to various lattice depths of interest and apply the adiabatic scaling law $T_r \propto \sqrt{U}$.
Finite radial temperatures lower the average trap depth experienced by the atoms from $U$ to an effective trap depth $U_{\mathrm{eff}}$.
We use
\begin{equation}
U_{\mathrm{eff}}=\int_0^{U} \rho(U^{'})U^{'}dU^{'}=U(1+k_BT_r/U)^{-1},
\label{eqn:eff}
\end{equation} 
 for its simplicity in tunneling rate calculations, where $\rho(U^{'})=\frac{1}{k_BT_r} \left(\frac{U^{'}}{U}\right)^{\frac{U}{k_BT_r}-1}$ is the probability density with respect to the local trap depth experienced by the atom and $k_B$ is the Boltzmann constant \cite{Ushijima2018, Beloy2020}.

\begin{figure}
    \centering
    \includegraphics[width=0.5\textwidth,height=170pt]{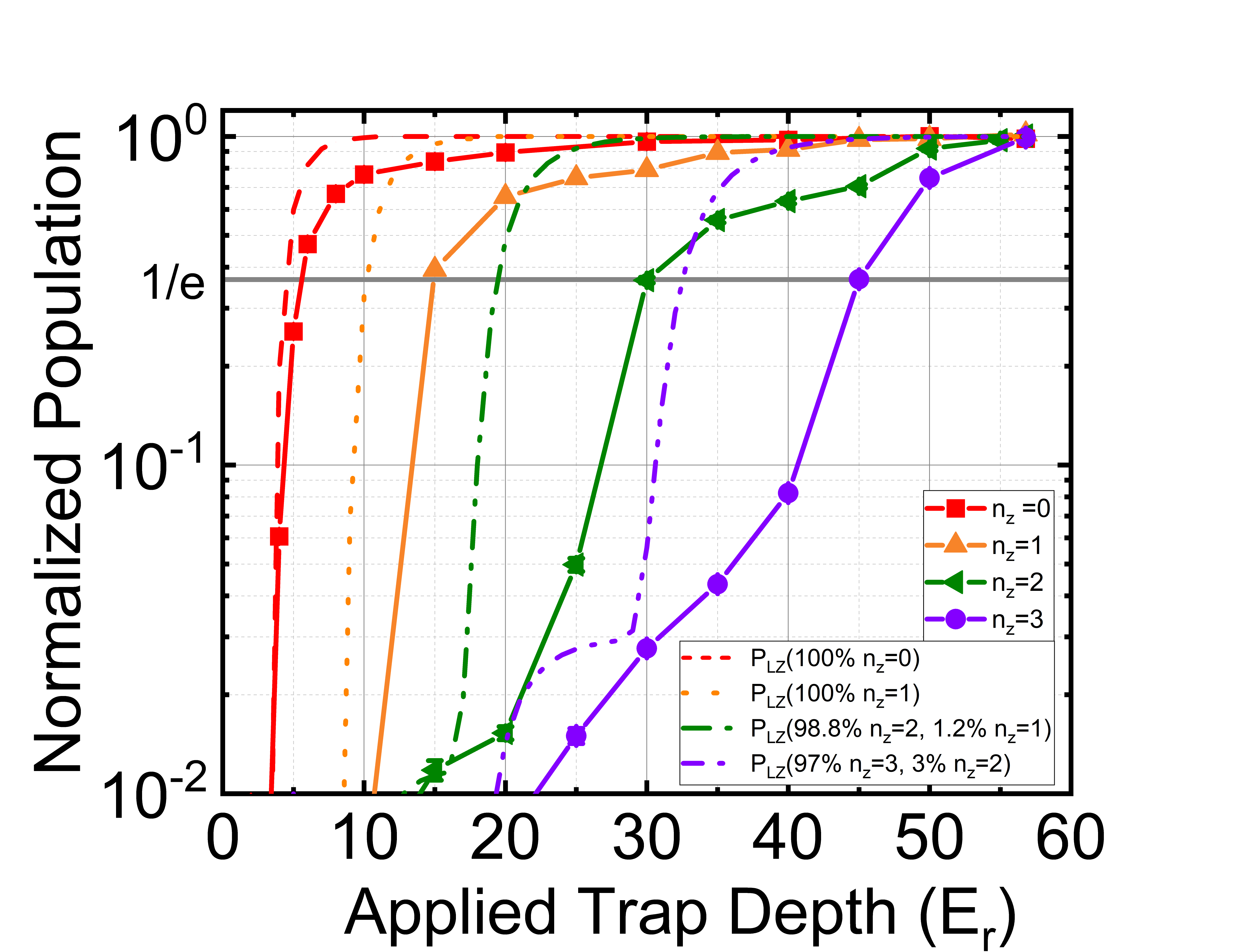}
    \caption{The population of atoms remaining after ramping to various applied trap depths from 57(1) $E_r$.
    Atoms are prepared in different longitudinal bands using ARP, with the targeted lattice band labeled as $n_z$. 
    Dashed lines are theoretically determined populations, $P_{LZ}(U,n_z)$, which show LZ tunneling leads to loss as the depth is lowered.
    The slight persistence in percent-level survival rates at low trap depths seen in $n_z=2,3$ is due to the percent level impurity in the targeted lattice band.
  }
    \label{fig:ARP_Purity}
\end{figure}

We prepare atoms in longitudinal bands ranging from $n_z=$ 0 to 3 (see appendix), adiabatically ramp to various trap depths, hold for $t_\mathrm{hold}=100$ ms, and adiabatically ramp back to a depth of 57 $E_r$.
The fraction of atoms remaining is plotted in Fig.~\ref{fig:ARP_Purity} against $U$. 
Based on linear interpolation of the data in Fig.~\ref{fig:ARP_Purity}, the applied trap depth at which a $1/e$ fraction remains is 5.5(6) $E_r$ for $n_z=0$ and 30.1(5) $E_r$ for $n_z=2$.
These depths are also theoretically calculated from the LZ tunneling rate
\begin{equation}
    R_{LZ}(U,n_z) \approx \nu_B e^{-\pi^2\Delta E(U,n_z)^2/(8mgE_rd)},
    \label{eqn:LZprob}
\end{equation}
 where $\Delta E(U,n_z)$ is the band gap between $n_z$ and $n_z+1$, $d=\lambda/2$, and $g$ is the acceleration due to gravity \cite{Anderson1998}. 
To better account for effective trap depth effects from the radial temperature, we compute an average fraction of atoms remaining
$P_{LZ}(U,n_z)=\int_0^{U} \rho(U^{'})\textrm{exp}[-t_{\mathrm{hold}} R_{LZ}(U^{'},n_z)]dU^{'}$, which are displayed as dashed lines on Fig.~\ref{fig:ARP_Purity} for $t_\mathrm{hold}=100$ ms.
We note that the time dependence and anharmonic nature of radial oscillations are not considered in $P_{LZ}$.

\begin{figure}
    \centering
    \includegraphics[width=0.5\textwidth]{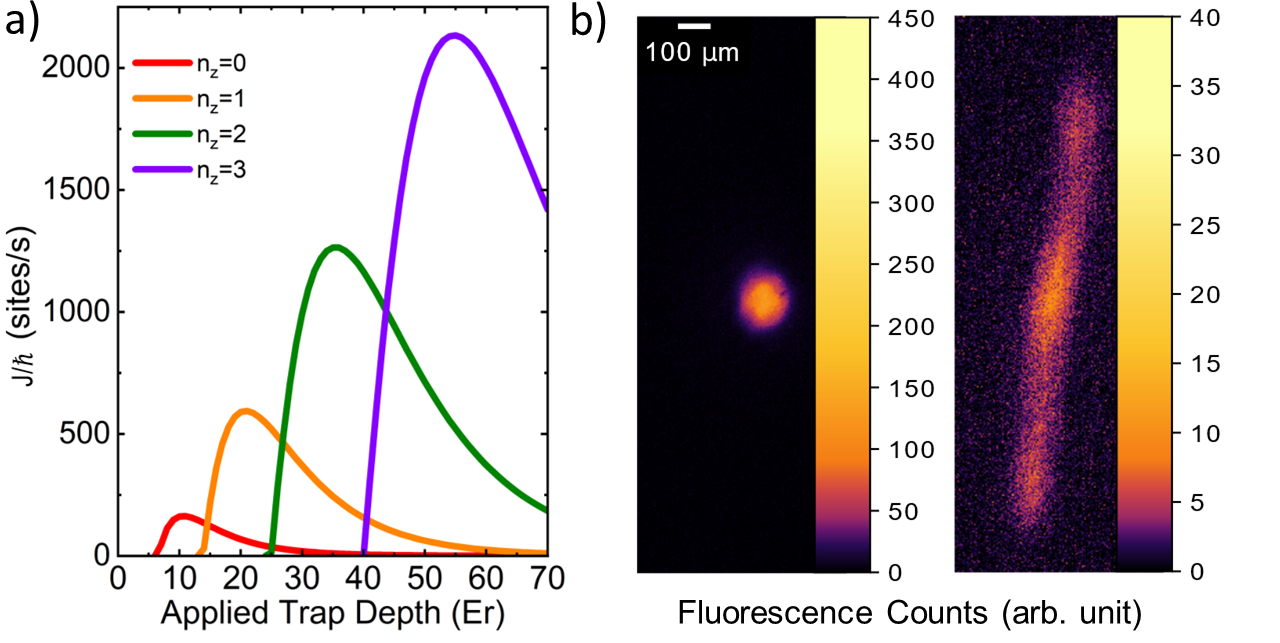}
    \caption{ a) We numerically integrate equation (\ref{eqn:tunnel}) to find the theoretical maximum tunneling rates in each motional band.
    Note the presence of effective trap depth scaling in the tunneling rate.
    We assume that the trap depth at any point during AM must not be lower than the cutoff depth, where $P_{LZ}(U,n_z)=1/e$.
    For a given band, this sets a maximum $\alpha$ for depths above the cutoff depth.
    b) Averaged fluorescence images of the delocalized Yb sample (right), and original sample (left). 
    Delocalization is applied at the experimentally determined optimal conditions for one second. 
    The tilt is imperfect alignment of the camera's vertical axis to the lattice axis and makes a negligible contribution to the determined size of the delocalized sample. 
     }
    \label{fig:TunnelTheory}
\end{figure}

Armed with measurements of LZ tunneling atom loss, we now consider the theoretically optimal conditions for coherent delocalization. 
For amplitude modulation at $\nu_B$, the nearest neighbor tunneling rate (in the single-band approximation) is
\begin{equation}
       J/\hbar= \frac{\alpha U_{\mathrm{eff}}}{2\hbar} \bra{\ell+1}\cos(2k_lz)\ket{\ell},
       \label{eqn:tunnel}
\end{equation}
where $\ket{\ell}$ is the WS wave function centered at site $\ell$, $z$ is the distance along the lattice, $\alpha$ is the AM depth, and $U_{\mathrm{eff}}$ is computed via equation (\ref{eqn:eff}) \cite{Alberti2010, Supplemental}.
The WS wave functions are numerically calculated for various trap depths and longitudinal motional bands.
At each trap depth we constrain $\alpha$ such that the lowest applied trap depth reached during modulation corresponds to $P_{LZ}(U,n_z)=1/e$ for $t_\mathrm{hold}=100$ ms \cite{Supplemental}.
The theoretical $J/\hbar$ plotted in Fig.~\ref{fig:TunnelTheory}(a) show that, for a constant minimum LZ lifetime, higher motional bands generally offer larger tunneling rates.
Alternatively, using the 1/$e$ measured loss thresholds from Fig.~\ref{fig:ARP_Purity} to constrain $\alpha$ also displays maximum tunneling rates that increase with $n_z$.

We experimentally measure delocalization with fluorescence imaging of the lattice-trapped, ultracold atoms. 
We elect to prepare atoms in $n_z=2$ with high purity using ARP and image the sample using 399-nm fluorescence.
Images are seen in Fig.~\ref{fig:TunnelTheory}(b) before and after coherent delocalization in $\bar{n}_z=$2.00(3), $U=40.0(5)$ $E_r$  (effective trap depth of 36.4(4) $E_r$), and $\alpha=0.3$  (the experimentally feasible fastest tunneling parameters). 
This coherent delocalization results in a modest $\sim$30\% atom loss due to LZ tunneling, in addition to 17(1)\% loss due to the ARP process. 
The left image is before tunneling, where the spatial extent of the atomic sample is set by the last stage of the 556-nm MOT. 
The right image shows, after coherent tunneling, that the full-width-half-maximum is approximately 7$\times$ larger, corresponding to a tunneling rate of $\sim 1800$ sites/s.
As shown in Fig.~\ref{fig:TunnelTheory}(a), lattice bands greater than $n_z=2$ could offer higher tunneling rates still.
Furthermore, Fig.~\ref{fig:TunnelTheory}(a) shows that, for approximately equal LZ losses, at optimal conditions $n_z=0$ takes $\sim8\times$ longer to reach an identical decrease in density when compared to $n_z=2$.

\begin{figure}
    \centering
    \includegraphics[width=0.5\textwidth]{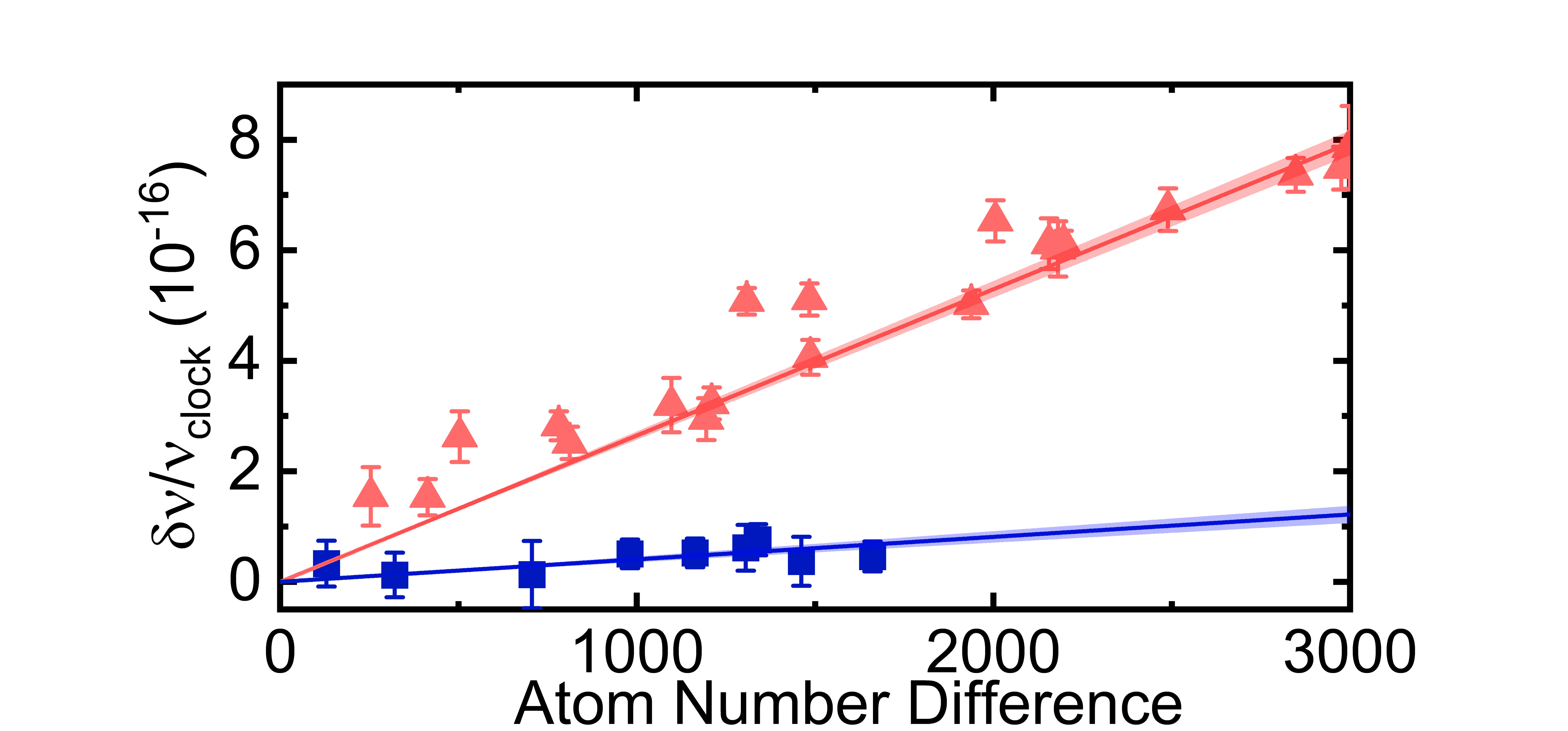}
    \caption{We measure the shift of the clock transition frequency between two non-spin-polarized samples of different atom numbers, with delocalized in blue squares and control (no delocalization) in red triangles. 
    For the control sample, data was taken at atom number differences larger than 3000, which are not plotted but still contribute to the fit.
    The red (blue) line gives a linear fit to the control (delocalized) measurements and shaded regions are 1-$\sigma$ statistical uncertainty. 
    Atom number is calibrated through fluorescence measurements, and all measurements are taken between 55 $E_r$ and 62 $E_r$. }
    \label{fig:density Shift}
\end{figure}

To highlight the benefit of delocalization, we measure the density dependent shift of the clock frequency in delocalized samples and compare to control samples without coherent delocalization.
To quantify the reduction in shift, for both the control case and delocalized test case, we forgo optical pumping to enhance the collisional shift effect (see appendix) and measure the frequency difference between two distinct numbers of atoms.
The shift versus the difference in atom number is plotted in Fig.~\ref{fig:density Shift}.
The error bars are the total Allan deviation at half the run length, with run lengths typically 1.5 hours long. 
A linear fit to the control data (red line) shows the shift is  $2.64(7)\times 10^{-19}$ per atom, in reasonable agreement with our previous measurements under somewhat different conditions \cite{McGrew2018}. 
In this case, some scatter in the shift can be seen, showcasing how day-to-day variations in experimental conditions may contribute to fluctuations in the observed collision shift.  
This underscores the utility of reducing the shift by means of a robust technique such as lower atomic density.
A linear fit to the delocalized data (blue line) shows a slope 6.5(8) times smaller than the control case. 
We expect this reduction in shift is entirely compatible with spin-polarized atomic samples (see appendix) or other density shift reduction techniques, including the larger lattice waists common in enhancement cavities \cite{LeTargat2013,McGrew2018,Bothwell2021}.

%This atom loss is directly at odds with the QPN goals of long spectroscopy at high atom numbers. 

\begin{figure}
    \centering
    \includegraphics[width=0.5\textwidth]{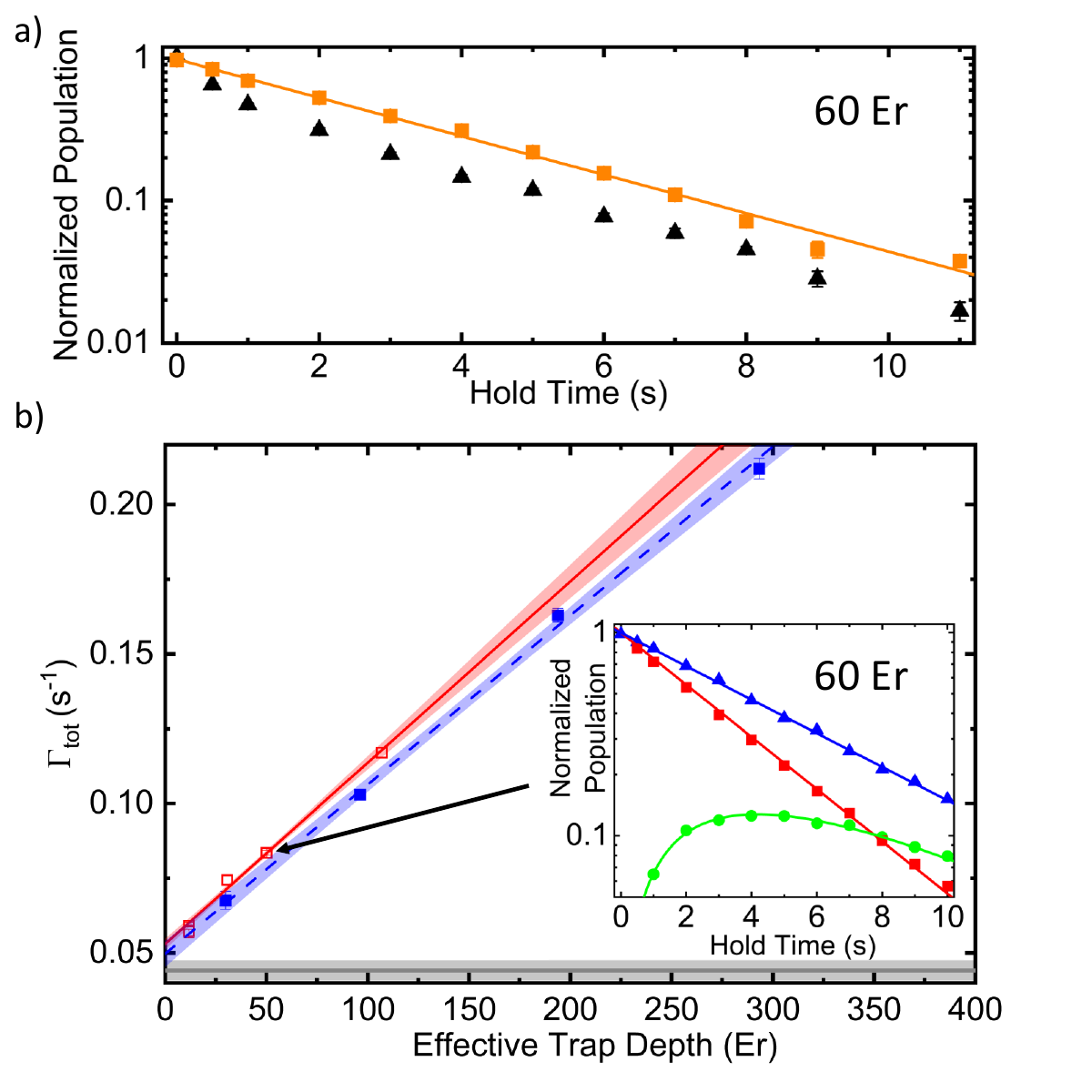}
    \caption{a) We measure the reduction in two-body loss from delocalization at $U=60(1)$ $E_r$. 
    Black triangles are $n_e(t)$  in non-delocalized samples, orange squares are $n_e(t)$ in delocalized samples, and the orange line is the fit to equation (\ref{eqn:ESdecay1}) for the delocalized sample. 
    Both samples have identical atom numbers to begin.
    b) Excited-state total decay rates are plotted versus the effective trap depth.
    Hollow red (solid blue) points are taken on different apparatus, and the solid red (dashed blue) line indicates the linear fit to each data set.  
    The gray line is a previous measurement of the clock state natural lifetime \cite{Xu2014}.
    Excited-state total decay rates are computed from fits such as the representative one in the inset, which was taken at $U$=59.8(5) $E_r$, corresponding to $U_{\mathrm{eff}}$=50.1 $E_r$. 
    For atoms beginning in the ground state (${}^1\textrm{S}_0$), $n_g$ are in blue triangles, and the blue line is a fit to $n_g(t)=n_g(t=0)\exp{(-\Gamma_{\textrm{loss}}t)}$.
    For atoms beginning in the excited ${}^3\textrm{P}_0$ state, $n_e$ ($n_g$) are plotted in red squares (green circles), and the red (green) line is the fit to equation (\ref{eqn:ESdecay1}) (equation (\ref{eqn:ESdecay2})), with $\Gamma_{\textrm{loss}}$ shared among all three fits for a given trap depth.
    The shaded areas are the 1-$\sigma$ statistical uncertainty regions.}
    \label{fig:ES_decay}
\end{figure}
In addition to reducing density-dependent systematic effects, coherent delocalization can reduce two-body loss for the benefit of clock stability.
Two-body loss originates from on-site inelastic collisions involving at least one atom in an excited electronic state \cite{Ludlow2011} and can degrade spectroscopic contrast at high densities or long spectroscopy times. 
To highlight the reduction in two-body loss from delocalization, we experimentally measure population loss in the $^1$S$_0$ and $^3$P$_0$ states.
In the absence of two-body loss the time-dependent populations ($n_g$ and $n_e$, respectively) are described by
\begin{equation}
    \dot{n_e}(t)=-\Gamma^{'}_{\textrm{loss}}n_e(t) -(\Gamma_0 +\gamma_L U_{\mathrm{eff}})n_e(t)
    \label{eqn:ESdecay1}
\end{equation}
\begin{equation}
    \dot{n_g}(t)=-\Gamma_{\textrm{loss}}n_g(t) +(\Gamma_0 +\gamma_L U_{\mathrm{eff}})n_e(t),
    \label{eqn:ESdecay2}
\end{equation}
which includes losses from the ground ($\Gamma_{\textrm{loss}}$) and excited ($\Gamma_{\textrm{loss}}^{'}$) states dominated by background gas collisions, spontaneous decay rate from the excited-state ($\Gamma_0$), and Raman-scattering-induced quenching of the excited-state proportional to the effective lattice depth ($\gamma_LU_{\mathrm{eff}}$). 
We note that the model deliberately does not include two-body loss mechanisms, which could induce non-exponential decay not seen in the model. 
For this model and all subsequently mentioned fits, we use a more careful treatment to calculate $U_{\mathrm{eff}}$ \cite{Beloy2020}, and also include negligible running wave effects.
We prepare non-spin-polarized atoms in the excited-state via ARP, blow away any remaining ground state atoms using light resonant with the 399-nm ${}^1\textrm{S}_0 \rightarrow {}^1\textrm{P}_1$ transition, hold for a variable time with no applied magnetic field, and finally measure the excited and ground state populations.
The populations are normalized to the number of atoms using an interleaved cycle employing no hold time, which reduces the effects of trapped atom number drifts over time. 
Each measurement is averaged for 70 experimental cycles or more.
Figure.~\ref{fig:ES_decay}(a) shows, in black triangles, the results for the excited-state population in a nondelocalized sample.
We observe prominent non-exponential behavior from two-body loss for hold times below two seconds.
After about two seconds, atom loss has decreased the density to the point where two-body loss is small and the remaining exponential loss is dominated by background gas collisions. 
Repeating an identical measurement with delocalized atoms, shown in orange squares, the solutions of equations (\ref{eqn:ESdecay1}) and (\ref{eqn:ESdecay2}) are fit to the normalized populations \footnote{Analytical solutions can be found at \cite{Dorscher2018}}.  
We see an excellent fit to the coupled differential equation model (orange line, reduced chi-squared statistic of 1.86) for an identical number of atoms, indicating negligible two-body loss.
Benefiting from the suppressed loss, at 3 seconds hold time the delocalized sample has 1.85(8) times more excited-state atoms remaining than the non-delocalized sample.

Without the nuisance of two-body loss, we can more easily study excited-state decay from lattice quenching ($\gamma_LU_{\mathrm{eff}}$) and spontaneous decay ($\Gamma_0$).
Lattice quenching deserves special attention: the magic wavelength is only 64 THz detuned from the ${}^3\textrm{P}_0 \rightarrow {}^3\textrm{S}_1$ E1 transition, leading to Raman scattering among the ${}^3\textrm{P}$ manifold.
%Raman scattering from ${}^3\textrm{P}_0$ to ${}^3\textrm{P}_2$ may lead to differences between $\Gamma_{\textrm{loss}}$ and $\Gamma_{\textrm{loss}}^{'}$; thus, they are independent parameters in all of our analyses.  
Raman scattering from ${}^3\textrm{P}_0$ to ${}^3\textrm{P}_1$ and the subsequent spontaneous emission to $^1$S$_0$ leads to a quenching rate of the clock transition ($\gamma_LU_{\mathrm{eff}}$) scaling linearly in $U_{\mathrm{eff}}$ \cite{Dorscher2018,Hutson2019}.

To quantify the effect, we measure $n_e$ and $n_g$ for atoms initially prepared in the excited-state over a range of hold times and trap depths. 
Spin-polarized and delocalized atoms are prepared identically to the above-described two-body loss measurement, and at each trap depth  equations (\ref{eqn:ESdecay1}) and (\ref{eqn:ESdecay2}) are fit to the populations. 
Representative data sets, taken at $U=59.8(5)~E_r$, along with their fits are shown in the inset of Fig.~\ref{fig:ES_decay}(b).
We note that, at every trap depth, $\Gamma_{\textrm{loss}}$ is simultaneously fit to a second data set of atoms prepared only in the ground state and susceptible only to losses from $\Gamma_{\textrm{loss}}$.
%The second data set's time dependent ground state population is fit to $\dot{n}_g(t)=-\Gamma_{\textrm{loss}} n_g(t)$.
We find $\Gamma_{\textrm{loss}}$ to be between $1.77(2)\times10^{-1}$ s\textsuperscript{$-1$} and $1.97(1)\times10^{-1}$ s\textsuperscript{$-1$}, depending on the date the data was taken on, and with no clear dependence on trap depth. 
$\Gamma^{'}_{\textrm{loss}}$ scales as $\approx \Gamma_{\textrm{loss}}(0.98+0.003\times U_{\mathrm{eff}}/E_r)$, with the trap depth dependent loss rate found to be $3.3(5) \times 10^{-4}$ $E_r$\textsuperscript{$-1$}s\textsuperscript{$-1$} \cite{Supplemental}.
Such linear scaling with $U_{\mathrm{eff}}$ is expected from Raman scattering to the untrapped ${}^3\textrm{P}_2$ state, theoretically predicted to be $3.5 \times 10^{-4}$ $E_r$\textsuperscript{$-1$}s\textsuperscript{$-1$} \cite{Supplemental}.
The total decay rate to the ground state $\Gamma_{tot}=\Gamma_0 + \gamma_L U_{\mathrm{eff}}$ is plotted against effective trap depth in Fig.~\ref{fig:ES_decay}(b) using hollow squares.

To verify the result, we conducted a similar measurement on a second distinct Yb lattice clock apparatus, shown as solid squares in Fig.~\ref{fig:ES_decay}.
While that system did not employ coherent delocalization, it uses a lattice enhancement cavity with a large waist \cite{McGrew2018}. 
Coupled with measurements limited to low atom number, we measured negligible two-body losses. 
This system did not utilize ARP, but rather a strong resonant drive on the clock transition to populate $^3$P$_0$.
For some data points radial cooling was used, while for others no radial cooling was used, consequently remaining more sensitive to the effective trap depth scaling. 
By fitting to a line, we find the quenching rate of the 759-nm lattice and the clock state natural lifetime for each independent clock apparatus, with agreement between apparatus at the 7\% level. 
We report a lattice quenching rate of $\gamma_L=5.7(7)\times10^{-4}$ $E_r$\textsuperscript{$-1$}s\textsuperscript{$-1$} based on the weighted mean of quenching rates for each clock apparatus, with weights of the inverse scatter in $\gamma_L$ when using different well-motivated methods of determining $U_{\mathrm{eff}}$ \cite{Ushijima2018, Beloy2020, Dorscher2018}, and the uncertainty taken as one-half the maximum difference between all methods on both clock apparatus. 
This quenching rate is in reasonable agreement with the theoretically predicted value \cite{Supplemental}.
The natural lifetime of 19(2) s is determined by an identical method, and is in 1.3-$\sigma$ agreement with a previously reported value \cite{Xu2014}.
Blackbody radiation decay from ${}^3\textrm{P}_0$ has a negligible effect on $\Gamma_0$: 
for 300 K operation, we theoretically compute the pumping rates of the dominant E1 transition (${}^3\textrm{P}_0$$\rightarrow$${}^3\textrm{D}_1$) and M1 transition (${}^3\textrm{P}_0\rightarrow{}^3\textrm{P}_1$) as $5.8\times10^{-9}~$s\textsuperscript{-1} and $6.7\times10^{-4}~$s\textsuperscript{-1}, respectively.

Reducing atomic density by means of coherent delocalization can enhance both systematic frequency shift and QPN-limited stability performance for the next generation of optical clocks.
Systematic frequency shifts due to cold collisions were reduced by a factor of 6.5(8) times. 
QPN is also lessened by reason of reduced two-body loss, in this case doubling the number of atoms remaining in the excited-clock-state for long interrogations enabled by the current generation of state-of-the-art cryogenic cavities \cite{Norcia2019}. 
Exploiting suppressed two-body loss, we measure the rate of Raman-scattering-induced quenching from the lattice and the excited-clock-state natural lifetime. 
We note that lattice quenching will unavoidably generate distinguishable unpolarized atoms, further increasing the need for density shift reduction techniques. 

The quantum control techniques presented here have applications beyond reducing density-dependent effects. 
State preparation in higher-lying $n_z$ bands is useful for determining the M1+E2 shift in optical lattice clocks \cite{Ushijima2018, Nemitz2019, Kim2022, Dorscher2022}. 
Clock transition ARP can remove the need for repump lasers in novel systems where repumping the excited-clock-states may not be feasible (see appendix) \cite{Witkowski2022, Fedorova2020}. 
Control of the tunneling rate can allow for quantum simulation of problems in complexity theory \cite{Muraleedharan2019} and for realization of Hamiltonians in tweezer arrays \cite{Spar2022}. 
Additionally, coherent delocalization can be a useful tool for clocks that spatially resolve their atomic samples \cite{Zheng2022, Bothwell2021}.

\begin{acknowledgments}
We gratefully acknowledge J. Lilieholm and T. Bothwell for careful reading of the manuscript. This work was supported by NIST, ONR, and NSF QLCI Award No. 2016244. 
\end{acknowledgments}

\appendix
\section{\label{app:ARP}Appendix A: Adiabatic Rapid Passage}
We efficiently prepare atoms in $n_z=$ 1, 2, and 3 using ARP on the clock (${}^1\textrm{S}_0 -{}^3\textrm{P}_0$) transition. 
The clock laser is co-linear with the optical lattice and benefits from resolved motional sidebands \cite{Blatt2009}. 
 At an operation depth of 60 $E_r$ and $T_r\sim 450$ nK, the clock laser's frequency is swept over the desired spectral feature ($\Delta n_z= -1,0,$ or $+1$) using a 14 kHz sweep range in 2 ms, while the intensity is modulated with an approximately Blackman profile.
 The peak carrier ($\ket{{}^1\textrm{S}_0, n_z=j} \rightarrow \ket{{}^3\textrm{P}_0, n_z=j}$) Rabi frequency is $\approx 14$ kHz, and the peak 1st order motional sideband ($\ket{{}^1\textrm{S}_0, n_z=j} \rightarrow \ket{{}^3\textrm{P}_0, n_z=j\pm1}$) Rabi frequency is $\approx 3.5$ kHz.
 The sweep range is chosen to maintain sufficient detuning from nearby motional sidebands when at $\sim$ 57 $E_r$, though the sweep range could be extended at larger trap depths (as could state preparation into higher $n_z$).
For the carrier transition we realize a transfer efficiency of 98(1)$\%$.

\begin{figure}
    \centering
    \includegraphics[width=0.5\textwidth]{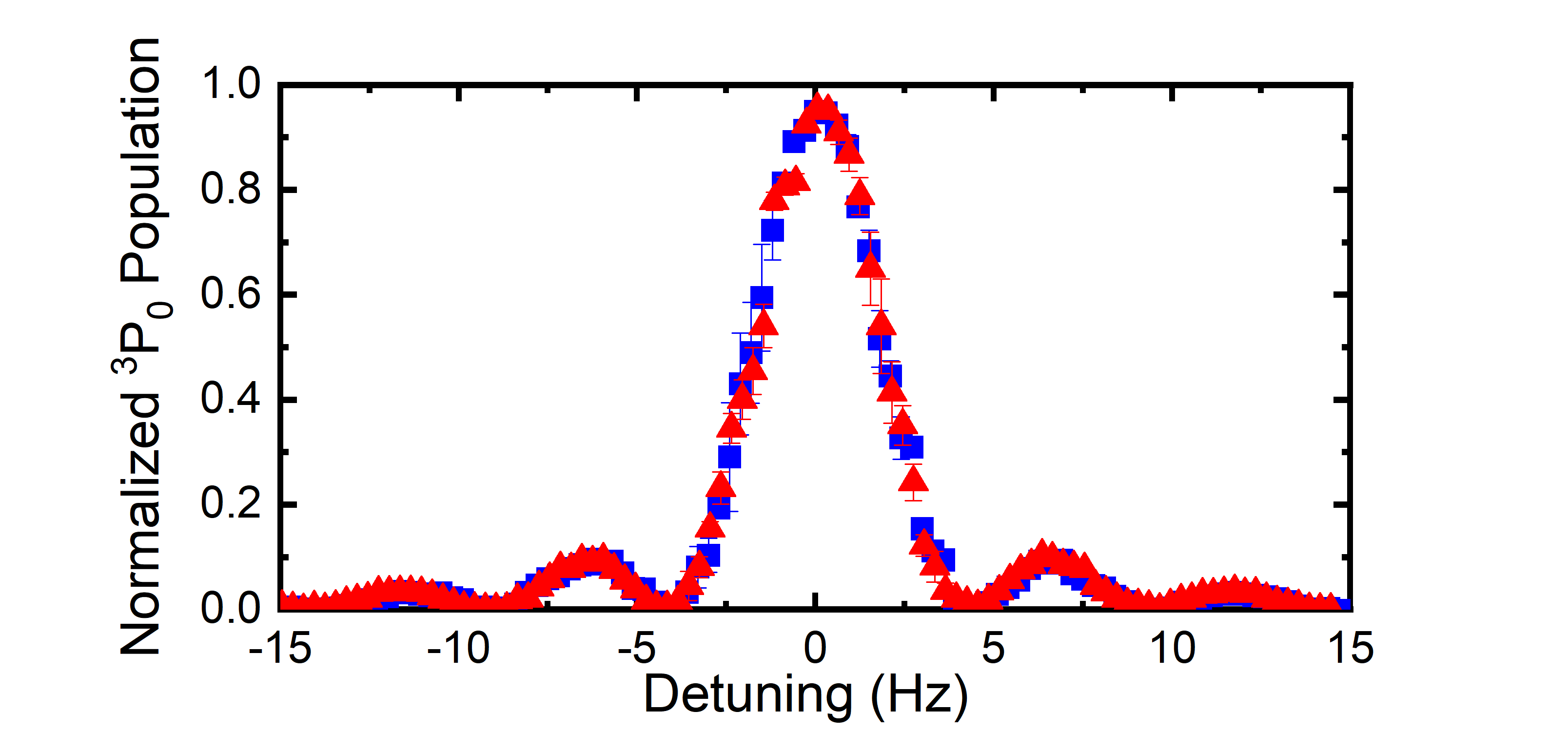}
    \caption{We measure a Rabi line by sweeping the frequency of our 578-nm laser over the clock transition. 
    The normalized population is measured using electron shelving, with ${}^{1}\textrm{S}_0$ atoms first measured using 399-nm fluorescence and ${}^3\textrm{P}_0$ atoms then measured by transferring them to ${}^1\textrm{S}_0$ then measuring 399-nm fluorescence. 
    We move atoms via the normal repump method (${}^3\textrm{P}_0 - {}^3\textrm{D}_1$ 1388-nm laser), shown in red triangles, or by ARP on the carrier transition, shown in blue squares. 
   }
    \label{fig:Rabi}
\end{figure}

The high ARP transfer efficiency on the carrier transition suggests that ARP would be a useful technique for $^3$P$_0$ state detection by means of applying ARP to transfer atoms to the ground state, and then cycling the $^1$S$_0$-$^1$P$_1$ transition.
  In most lattice clocks, it is common to use one or more optical pumping lasers to $^3$D$_1$ or $^3$S$_1$ for this function, at the cost of additional laser wavelengths and lossy decay channels.  
  To illustrate the utility of ARP for state detection, we demonstrate narrow-line Rabi spectroscopy in Fig.~\ref{fig:Rabi} using both ARP and a more traditional 1388-nm $^3$P$_0$-$^3$D$_1$ optical pumping laser.

To realize our choice of $n_z=2$ for coherent delocalization, we first prepare the atoms in the ground motional band using Sisyphus cooling \cite{Chen2022}.
 To prepare atoms in $n_z=2$ for delocalization, we first transfer the initial $\ket{{}^1\textrm{S}_0, n_z=0}$ population to $\ket{{}^3\textrm{P}_0, n_z=1}$ via ARP.
 The $\Delta n_z=1$ sideband corner frequency is 30 kHz at a 60 $E_r$ trap depth, and we choose a central ARP frequency of 24.5 kHz. 
 The $\Delta n_z=1$ ARP has identical parameters to the carrier ARP: a sweep time of 2 ms, a 14 kHz sweep range, intensity modulated with an approximately Blackman profile, and an identical peak intensity corresponding to a carrier Rabi frequency of $\sim 14$ kHz, which is a $\Delta n_z=1$ sideband Rabi frequency of $\sim 3.5$ kHz. 
 This sideband ARP realizes a transfer efficiency to $\ket{{}^3\textrm{P}_0, n_z=1}$ of 90(1)\%. 
 The remaining impurity, $\ket{{}^1\textrm{S}_0, n_z=0}$, is heated out of the lattice using 399-nm light resonant with the ${}^1\textrm{S}_0 - {}^1\textrm{P}_1$ transition. 
 We then apply ARP on the carrier to move population to $\ket{{}^1\textrm{S}_0, n_z=1}$, subsequently repumping any small fraction of atoms remaining in ${}^3\textrm{P}_0$ using 1388-nm light. 
 (The ARP on the carrier, compared to 1388-nm repumping, ensures as few atoms as possible experience a spontaneous emission, which can induce unwanted changes in $n_z$ populations, thereby maximizing motional state purity.)
 This process is applied $j$ times to obtain atoms in $\ket{{}^1\textrm{S}_0, n_z=j}$.
 For $j$=2, the overall process is 83(1)\% efficient, and atom losses are predominantly the result of the 399-nm excitation to keep the sample as pure as possible. 
  For $j>2$, trap anharmonicity substantially changes the corner frequency of the sideband, resulting in reduced transfer efficiency, though this could be improved by dynamically modifying the center frequency of the ARP. 
  
 To measure the preparation purity of $\ket{n_z=j}$, we adiabatically ramp (0.6 $E_r$/ms) a sample prepared in $\ket{n_z=j}$ to a trap depth for $t_\mathrm{hold}=100$ ms, which we computed to have a $P_{LZ}(U,n_z=j) \approx 0.005$. 
 We assume that the remaining population is the impurity $\ket{n_z<j}$.
 To account for LZ tunneling of $\ket{n_z<j}$, we take atoms prepared in $\ket{n_z=j-1}$, subject them to the same adiabatic ramp, and normalize the impurity by the fraction remaining.
 For our sample prepared in $n_z=2$, we measure an $n_z<2$ impurity of 3.0(6)\%.
 
 To de-excite atoms back to $\ket{{}^1\textrm{S}_0, n_z=0}$ after coherent delocalization, we twice perform: ARP on the red sideband, ARP on the carrier, and then repumping. 
 From here, we de-excite the small residual $\ket{n_z>0}$ population by thrice applying: ARP on the red sideband followed by repumping. 
 Using longitudinal sideband spectroscopy, we measure the final longitudinal temperature to be $\sim0.5$ $\mu$K, which is slightly colder than was achieved directly from Sisyphus cooling. 
 The entire ARP and delocalization process has also been measured, using longitudinal sideband spectroscopy, to decrease the radial temperature by $<$ 200(100) nK; therefore, we conclude that ARP and delocalization does not lead to detrimental heating.
 We also demonstrated preparation of $n_z=2$ atoms by performing ARP on the second-order longitudinal sideband ($\ket{{}^1\textrm{S}_0, n_z=0} \rightarrow \ket{{}^3\textrm{P}_0, n_z=2}$), but found that the above-described protocol could realize the desired $n_z=2$ sample with less loss and greater purity.

\section{Appendix B: Imaging}
To measure the tunneling rate, we first adiabatically ramp the applied trap depth from $\sim$ 57 $E_r$, then apply one second of AM to the voltage reference of our trap light intensity servo that acts on an acoustic-optic modulator, and finally adiabatically ramp back to $\sim$ 57 $E_r$ for imaging. 
During imaging we apply light resonant with the ${}^1\textrm{S}_0 - {}^1\textrm{P}_1$ transition and collect the fluorescence on a CMOS camera.

The corresponding tunneling rate of $\sim 1800$ sites/s can be calculated as
 \begin{equation}
     J/\hbar=  \frac{2\sqrt{2}}{\lambda}\sqrt{\frac{\sigma_f^2-\sigma_i^2}{t^2}},
 \end{equation}
 where $\sigma_{f(i)}$ is the final (initial) one dimensional vertical spread of the atomic sample from a Gaussian fit and $t$ is the time duration of amplitude modulation \cite{Alberti2010}.
The experimentally determined maximum tunneling rates agree with the qualitative $n_z$ scaling from theory, but we measure an excess beyond the predicted tunneling rates in all $n_z$ bands (the excess at the experimentally optimal delocalization condition is $\sim$ 75\%).

\section{Appendix C: Non-Spin-Polarized vs. Spin-Polarized Density Shifts}
Without optical pumping, our fermionic \textsuperscript{171}Yb atoms populate both $m_F=\pm 1/2$ Zeeman states, and $s$-wave collisional shifts between these are not suppressed by Pauli exclusion.
This yields a larger cold-collision shift than traditional, spin-polarized samples (where $p$-wave shifts usually dominate), deliberately chosen here to make measurement of the shift easier.
Though the shift is larger in absolute terms, since both $s$- and $p$-wave shifts remain linear in density for typical operational conditions, the degree of suppression we measure here is indicative of the degree of suppression that would be realized for $p$-wave shifts in a spin-polarized sample. 
\section{Appendix D: Coherent Tunneling} 
\begin{figure}
\centering
    \includegraphics[width=0.5\textwidth]{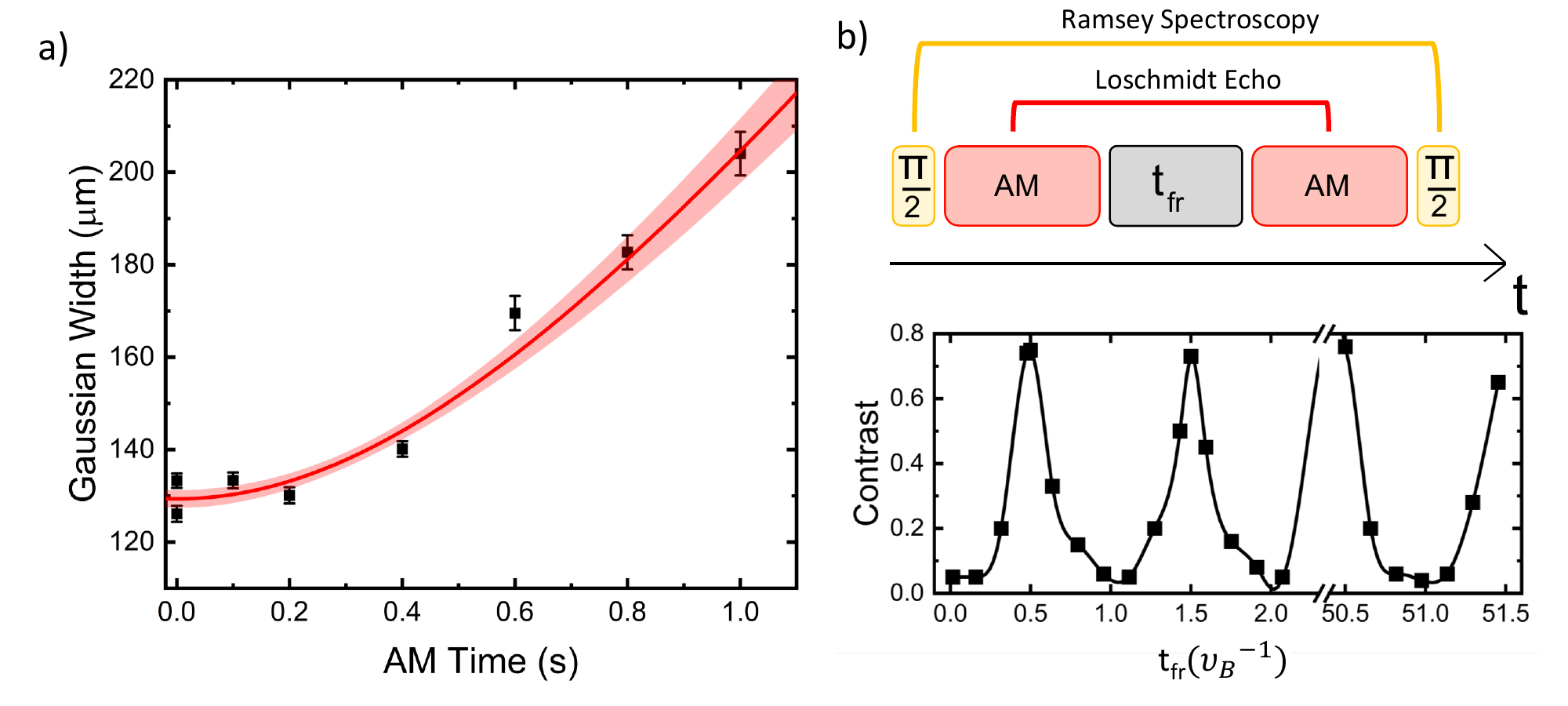}
    \caption{Coherence in tunneling is demonstrated. a) The vertical size of the atomic sample is plotted vs time. The red line shows a fit to the expected coherent tunneling dynamics of $\sqrt{\sigma_0+v^2t^2}$, with the shaded region at 1-$\sigma$ statistical uncertainty. The vertical size is measured via Gaussian fit. The tunneling parameters are atoms in $n_z=0$ at 10 $E_r$ with $\alpha=0.4$.
    b) Reversal of coherent tunneling is demonstrated. A Loschmidt echo (two pulses of coherent delocalization (AM) separated by a freezing time $t_{\mathrm{fr}}$) is performed during the dark time of Ramsey spectroscopy. 
    The contrast of the Ramsey lineshape is measured to have a period of the freezing time, indicating that the coherent tunneling dynamics can be reversed. 
    Lines are a guide to the eye.
    }
    \label{fig:coherent}
\end{figure}

The coherent tunneling demonstrated in this paper has numerous attractive features for site-to-site wavefunction manipulation. 
In $n_z=0$, we demonstrate a Fourier limited resonance at $\nu_B$ in measurements of the tunneling rate. 
Coherent tunneling has wavefunction size that increases $\propto t$ at the limit of long times, as opposed to $\propto \sqrt{t}$ for incoherent tunneling.
We observe size increases scaling as $\sim\sqrt{\sigma_0+v^2t^2}$ in Fig.~\ref{fig:coherent}(a), as expected for coherent delocalization \cite{Alberti2010}, where $\sigma_0$ is the initial size, and $v$ is the tunneling velocity.

We also demonstrate coherence via Loschmidt echoes \cite{Jalabert2001}. 
We separate two identical bursts of AM (individually realizing coherent delocalization) by a freezing time $t_{\mathrm{fr}}$, where no resonant tunneling takes place, but neighboring sites accumulate a phase difference of $2\pi\times\nu_B t_{\mathrm{fr}}$. 
Thus, the wavefunction size increase of the first AM burst can be reversed after the 2nd AM burst using the periodic response of the freezing time \cite{Alberti2010}. 
We probe this reversal of tunneling using contrast measurements in Ramsey spectroscopy. 
Due to the incommensurate nature of the Yb magic wavelength and clock transition wavelength, we make the assumption that tunneling over several lattice sites leads to a random accumulation of clock laser phase. 
Accordingly, we expect tunneling during Ramsey spectroscopy to lead to a reduction in contrast.
The Loschmidt echo sequence is performed between the two $\frac{\pi}{2}$ pulses of Ramsey spectroscopy, see Fig.~\ref{fig:coherent}(b). 
The contrast of the Ramsey lineshape is plotted vs. the freezing time of the Loschmidt echo sequence in Fig.~\ref{fig:coherent}(b).
We see clear periodicity in the freezing time, with period $1/\nu_{B}$, showing that phase accumulation from the first coherent tunneling pulse can be reversed by the second.
This indicates that we could reverse the size increase of the wavefunction using coherent tunneling dynamics.

\nocite{SuppCite}
\nocite{Lemonde2005}
\nocite{BLATT1967382}
\nocite{Zheng2023}
\nocite{Beloy2014}
\nocite{Beloy2018}
\nocite{SavJoh02}
\nocite{SafPorCla12}
\nocite{DzuDer10}
\nocite{Bel12}

\begin{thebibliography}{63}%
\makeatletter
\providecommand \@ifxundefined [1]{%
 \@ifx{#1\undefined}
}%
\providecommand \@ifnum [1]{%
 \ifnum #1\expandafter \@firstoftwo
 \else \expandafter \@secondoftwo
 \fi
}%
\providecommand \@ifx [1]{%
 \ifx #1\expandafter \@firstoftwo
 \else \expandafter \@secondoftwo
 \fi
}%
\providecommand \natexlab [1]{#1}%
\providecommand \enquote  [1]{``#1''}%
\providecommand \bibnamefont  [1]{#1}%
\providecommand \bibfnamefont [1]{#1}%
\providecommand \citenamefont [1]{#1}%
\providecommand \href@noop [0]{\@secondoftwo}%
\providecommand \href [0]{\begingroup \@sanitize@url \@href}%
\providecommand \@href[1]{\@@startlink{#1}\@@href}%
\providecommand \@@href[1]{\endgroup#1\@@endlink}%
\providecommand \@sanitize@url [0]{\catcode `\\12\catcode `\$12\catcode
  `\&12\catcode `\#12\catcode `\^12\catcode `\_12\catcode `\%12\relax}%
\providecommand \@@startlink[1]{}%
\providecommand \@@endlink[0]{}%
\providecommand \url  [0]{\begingroup\@sanitize@url \@url }%
\providecommand \@url [1]{\endgroup\@href {#1}{\urlprefix }}%
\providecommand \urlprefix  [0]{URL }%
\providecommand \Eprint [0]{\href }%
\providecommand \doibase [0]{https://doi.org/}%
\providecommand \selectlanguage [0]{\@gobble}%
\providecommand \bibinfo  [0]{\@secondoftwo}%
\providecommand \bibfield  [0]{\@secondoftwo}%
\providecommand \translation [1]{[#1]}%
\providecommand \BibitemOpen [0]{}%
\providecommand \bibitemStop [0]{}%
\providecommand \bibitemNoStop [0]{.\EOS\space}%
\providecommand \EOS [0]{\spacefactor3000\relax}%
\providecommand \BibitemShut  [1]{\csname bibitem#1\endcsname}%
\let\auto@bib@innerbib\@empty
%</preamble>
\bibitem [{\citenamefont {Ushijima}\ \emph {et~al.}(2015)\citenamefont
  {Ushijima}, \citenamefont {Takamoto}, \citenamefont {Das}, \citenamefont
  {Ohkubo},\ and\ \citenamefont {Katori}}]{Ushijima2015}%
  \BibitemOpen
  \bibfield  {author} {\bibinfo {author} {\bibfnamefont {I.}~\bibnamefont
  {Ushijima}}, \bibinfo {author} {\bibfnamefont {M.}~\bibnamefont {Takamoto}},
  \bibinfo {author} {\bibfnamefont {M.}~\bibnamefont {Das}}, \bibinfo {author}
  {\bibfnamefont {T.}~\bibnamefont {Ohkubo}},\ and\ \bibinfo {author}
  {\bibfnamefont {H.}~\bibnamefont {Katori}},\ }\bibfield  {title} {\bibinfo
  {title} {Cryogenic optical lattice clocks},\ }\href
  {https://doi.org/10.1038/nphoton.2015.5} {\bibfield  {journal} {\bibinfo
  {journal} {Nature Photonics}\ }\textbf {\bibinfo {volume} {9}},\ \bibinfo
  {pages} {185} (\bibinfo {year} {2015})}\BibitemShut {NoStop}%
\bibitem [{\citenamefont {McGrew}\ \emph {et~al.}(2018)\citenamefont {McGrew},
  \citenamefont {Zhang}, \citenamefont {Fasano}, \citenamefont {Sch{\"a}ffer},
  \citenamefont {Beloy}, \citenamefont {Nicolodi}, \citenamefont {Brown},
  \citenamefont {Hinkley}, \citenamefont {Milani}, \citenamefont {Schioppo},
  \citenamefont {Yoon},\ and\ \citenamefont {Ludlow}}]{McGrew2018}%
  \BibitemOpen
  \bibfield  {author} {\bibinfo {author} {\bibfnamefont {W.~F.}\ \bibnamefont
  {McGrew}}, \bibinfo {author} {\bibfnamefont {X.}~\bibnamefont {Zhang}},
  \bibinfo {author} {\bibfnamefont {R.~J.}\ \bibnamefont {Fasano}}, \bibinfo
  {author} {\bibfnamefont {S.~A.}\ \bibnamefont {Sch{\"a}ffer}}, \bibinfo
  {author} {\bibfnamefont {K.}~\bibnamefont {Beloy}}, \bibinfo {author}
  {\bibfnamefont {D.}~\bibnamefont {Nicolodi}}, \bibinfo {author}
  {\bibfnamefont {R.~C.}\ \bibnamefont {Brown}}, \bibinfo {author}
  {\bibfnamefont {N.}~\bibnamefont {Hinkley}}, \bibinfo {author} {\bibfnamefont
  {G.}~\bibnamefont {Milani}}, \bibinfo {author} {\bibfnamefont
  {M.}~\bibnamefont {Schioppo}}, \bibinfo {author} {\bibfnamefont {T.~H.}\
  \bibnamefont {Yoon}},\ and\ \bibinfo {author} {\bibfnamefont {A.~D.}\
  \bibnamefont {Ludlow}},\ }\bibfield  {title} {\bibinfo {title} {Atomic clock
  performance enabling geodesy below the centimetre level},\ }\href
  {https://doi.org/10.1038/s41586-018-0738-2} {\bibfield  {journal} {\bibinfo
  {journal} {Nature}\ }\textbf {\bibinfo {volume} {564}},\ \bibinfo {pages}
  {87} (\bibinfo {year} {2018})}\BibitemShut {NoStop}%
\bibitem [{\citenamefont {Bothwell}\ \emph {et~al.}(2019)\citenamefont
  {Bothwell}, \citenamefont {Kedar}, \citenamefont {Oelker}, \citenamefont
  {Robinson}, \citenamefont {Bromley}, \citenamefont {Tew}, \citenamefont
  {Ye},\ and\ \citenamefont {Kennedy}}]{Bothwell2019}%
  \BibitemOpen
  \bibfield  {author} {\bibinfo {author} {\bibfnamefont {T.}~\bibnamefont
  {Bothwell}}, \bibinfo {author} {\bibfnamefont {D.}~\bibnamefont {Kedar}},
  \bibinfo {author} {\bibfnamefont {E.}~\bibnamefont {Oelker}}, \bibinfo
  {author} {\bibfnamefont {J.~M.}\ \bibnamefont {Robinson}}, \bibinfo {author}
  {\bibfnamefont {S.~L.}\ \bibnamefont {Bromley}}, \bibinfo {author}
  {\bibfnamefont {W.~L.}\ \bibnamefont {Tew}}, \bibinfo {author} {\bibfnamefont
  {J.}~\bibnamefont {Ye}},\ and\ \bibinfo {author} {\bibfnamefont {C.~J.}\
  \bibnamefont {Kennedy}},\ }\bibfield  {title} {\bibinfo {title} {{{JILA}
  {SrI} optical lattice clock with uncertainty of $2.0 \times 10^{-18}$}},\
  }\href {https://doi.org/10.1088/1681-7575/ab4089} {\bibfield  {journal}
  {\bibinfo  {journal} {Metrologia}\ }\textbf {\bibinfo {volume} {56}},\
  \bibinfo {pages} {065004} (\bibinfo {year} {2019})}\BibitemShut {NoStop}%
\bibitem [{\citenamefont {Wcis{\l}o}\ \emph {et~al.}(2018)\citenamefont
  {Wcis{\l}o}, \citenamefont {Ablewski}, \citenamefont {Beloy}, \citenamefont
  {Bilicki}, \citenamefont {Bober}, \citenamefont {Brown}, \citenamefont
  {Fasano}, \citenamefont {Ciury{\l}o}, \citenamefont {Hachisu}, \citenamefont
  {Ido}, \citenamefont {Lodewyck}, \citenamefont {Ludlow}, \citenamefont
  {McGrew}, \citenamefont {Morzy{\'{n}}ski}, \citenamefont {Nicolodi},
  \citenamefont {Schioppo}, \citenamefont {Sekido}, \citenamefont {{Le
  Targat}}, \citenamefont {Wolf}, \citenamefont {Zhang}, \citenamefont
  {Zjawin},\ and\ \citenamefont {Zawada}}]{Wciso2018}%
  \BibitemOpen
  \bibfield  {author} {\bibinfo {author} {\bibfnamefont {P.}~\bibnamefont
  {Wcis{\l}o}}, \bibinfo {author} {\bibfnamefont {P.}~\bibnamefont {Ablewski}},
  \bibinfo {author} {\bibfnamefont {K.}~\bibnamefont {Beloy}}, \bibinfo
  {author} {\bibfnamefont {S.}~\bibnamefont {Bilicki}}, \bibinfo {author}
  {\bibfnamefont {M.}~\bibnamefont {Bober}}, \bibinfo {author} {\bibfnamefont
  {R.}~\bibnamefont {Brown}}, \bibinfo {author} {\bibfnamefont
  {R.}~\bibnamefont {Fasano}}, \bibinfo {author} {\bibfnamefont
  {R.}~\bibnamefont {Ciury{\l}o}}, \bibinfo {author} {\bibfnamefont
  {H.}~\bibnamefont {Hachisu}}, \bibinfo {author} {\bibfnamefont
  {T.}~\bibnamefont {Ido}}, \bibinfo {author} {\bibfnamefont {J.}~\bibnamefont
  {Lodewyck}}, \bibinfo {author} {\bibfnamefont {A.}~\bibnamefont {Ludlow}},
  \bibinfo {author} {\bibfnamefont {W.}~\bibnamefont {McGrew}}, \bibinfo
  {author} {\bibfnamefont {P.}~\bibnamefont {Morzy{\'{n}}ski}}, \bibinfo
  {author} {\bibfnamefont {D.}~\bibnamefont {Nicolodi}}, \bibinfo {author}
  {\bibfnamefont {M.}~\bibnamefont {Schioppo}}, \bibinfo {author}
  {\bibfnamefont {M.}~\bibnamefont {Sekido}}, \bibinfo {author} {\bibfnamefont
  {R.}~\bibnamefont {{Le Targat}}}, \bibinfo {author} {\bibfnamefont
  {P.}~\bibnamefont {Wolf}}, \bibinfo {author} {\bibfnamefont {X.}~\bibnamefont
  {Zhang}}, \bibinfo {author} {\bibfnamefont {B.}~\bibnamefont {Zjawin}},\ and\
  \bibinfo {author} {\bibfnamefont {M.}~\bibnamefont {Zawada}},\ }\bibfield
  {title} {\bibinfo {title} {{New bounds on dark matter coupling from a global
  network of optical atomic clocks}},\ }\href
  {https://doi.org/10.1126/sciadv.aau4869} {\bibfield  {journal} {\bibinfo
  {journal} {Science Advances}\ }\textbf {\bibinfo {volume} {4}},\ \bibinfo
  {pages} {eaau4869} (\bibinfo {year} {2018})}\BibitemShut {NoStop}%
\bibitem [{\citenamefont {Wcis{\l}o}\ \emph {et~al.}(2017)\citenamefont
  {Wcis{\l}o}, \citenamefont {Morzy{\'{n}}ski}, \citenamefont {Bober},
  \citenamefont {Cygan}, \citenamefont {Lisak}, \citenamefont {Ciury{\l}o},\
  and\ \citenamefont {Zawada}}]{Wciso2017}%
  \BibitemOpen
  \bibfield  {author} {\bibinfo {author} {\bibfnamefont {P.}~\bibnamefont
  {Wcis{\l}o}}, \bibinfo {author} {\bibfnamefont {P.}~\bibnamefont
  {Morzy{\'{n}}ski}}, \bibinfo {author} {\bibfnamefont {M.}~\bibnamefont
  {Bober}}, \bibinfo {author} {\bibfnamefont {A.}~\bibnamefont {Cygan}},
  \bibinfo {author} {\bibfnamefont {D.}~\bibnamefont {Lisak}}, \bibinfo
  {author} {\bibfnamefont {R.}~\bibnamefont {Ciury{\l}o}},\ and\ \bibinfo
  {author} {\bibfnamefont {M.}~\bibnamefont {Zawada}},\ }\bibfield  {title}
  {\bibinfo {title} {Experimental constraint on dark matter detection with
  optical atomic clocks},\ }\href {https://doi.org/10.1038/s41550-016-0009}
  {\bibfield  {journal} {\bibinfo  {journal} {Nature Astronomy}\ }\textbf
  {\bibinfo {volume} {1}},\ \bibinfo {pages} {0009} (\bibinfo {year}
  {2017})}\BibitemShut {NoStop}%
\bibitem [{\citenamefont {{Boulder Atomic Clock Optical Network (BACON)
  Collaboration*}}(2021)}]{Beloy2021}%
  \BibitemOpen
  \bibfield  {author} {\bibinfo {author} {\bibnamefont {{Boulder Atomic Clock
  Optical Network (BACON) Collaboration*}}},\ }\bibfield  {title} {\bibinfo
  {title} {Frequency ratio measurements at 18-digit accuracy using an optical
  clock network},\ }\href {https://doi.org/10.1038/s41586-021-03253-4}
  {\bibfield  {journal} {\bibinfo  {journal} {Nature}\ }\textbf {\bibinfo
  {volume} {591}},\ \bibinfo {pages} {564} (\bibinfo {year}
  {2021})}\BibitemShut {NoStop}%
\bibitem [{\citenamefont {Derevianko}\ and\ \citenamefont
  {Pospelov}(2014)}]{Derevianko2014}%
  \BibitemOpen
  \bibfield  {author} {\bibinfo {author} {\bibfnamefont {A.}~\bibnamefont
  {Derevianko}}\ and\ \bibinfo {author} {\bibfnamefont {M.}~\bibnamefont
  {Pospelov}},\ }\bibfield  {title} {\bibinfo {title} {Hunting for topological
  dark matter with atomic clocks},\ }\href {https://doi.org/10.1038/nphys3137}
  {\bibfield  {journal} {\bibinfo  {journal} {Nature Physics}\ }\textbf
  {\bibinfo {volume} {10}},\ \bibinfo {pages} {933} (\bibinfo {year}
  {2014})}\BibitemShut {NoStop}%
\bibitem [{\citenamefont {Chou}\ \emph {et~al.}(2010)\citenamefont {Chou},
  \citenamefont {Hume}, \citenamefont {Rosenband},\ and\ \citenamefont
  {Wineland}}]{Chou2010}%
  \BibitemOpen
  \bibfield  {author} {\bibinfo {author} {\bibfnamefont {C.~W.}\ \bibnamefont
  {Chou}}, \bibinfo {author} {\bibfnamefont {D.~B.}\ \bibnamefont {Hume}},
  \bibinfo {author} {\bibfnamefont {T.}~\bibnamefont {Rosenband}},\ and\
  \bibinfo {author} {\bibfnamefont {D.~J.}\ \bibnamefont {Wineland}},\
  }\bibfield  {title} {\bibinfo {title} {{Optical Clocks and Relativity}},\
  }\href {https://doi.org/10.1126/science.1192720} {\bibfield  {journal}
  {\bibinfo  {journal} {Science}\ }\textbf {\bibinfo {volume} {239}},\ \bibinfo
  {pages} {1630} (\bibinfo {year} {2010})}\BibitemShut {NoStop}%
\bibitem [{\citenamefont {Grotti}\ \emph {et~al.}(2018)\citenamefont {Grotti},
  \citenamefont {Koller}, \citenamefont {Vogt}, \citenamefont {H{\"a}fner},
  \citenamefont {Sterr}, \citenamefont {Lisdat}, \citenamefont {Denker},
  \citenamefont {Voigt}, \citenamefont {Timmen}, \citenamefont {Rolland},
  \citenamefont {Baynes}, \citenamefont {Margolis}, \citenamefont {Zampaolo},
  \citenamefont {Thoumany}, \citenamefont {Pizzocaro}, \citenamefont {Rauf},
  \citenamefont {Bregolin}, \citenamefont {Tampellini}, \citenamefont
  {Barbieri}, \citenamefont {Zucco}, \citenamefont {Costanzo}, \citenamefont
  {Clivati}, \citenamefont {Levi},\ and\ \citenamefont
  {Calonico}}]{Grotti2018}%
  \BibitemOpen
  \bibfield  {author} {\bibinfo {author} {\bibfnamefont {J.}~\bibnamefont
  {Grotti}}, \bibinfo {author} {\bibfnamefont {S.}~\bibnamefont {Koller}},
  \bibinfo {author} {\bibfnamefont {S.}~\bibnamefont {Vogt}}, \bibinfo {author}
  {\bibfnamefont {S.}~\bibnamefont {H{\"a}fner}}, \bibinfo {author}
  {\bibfnamefont {U.}~\bibnamefont {Sterr}}, \bibinfo {author} {\bibfnamefont
  {C.}~\bibnamefont {Lisdat}}, \bibinfo {author} {\bibfnamefont
  {H.}~\bibnamefont {Denker}}, \bibinfo {author} {\bibfnamefont
  {C.}~\bibnamefont {Voigt}}, \bibinfo {author} {\bibfnamefont
  {L.}~\bibnamefont {Timmen}}, \bibinfo {author} {\bibfnamefont
  {A.}~\bibnamefont {Rolland}}, \bibinfo {author} {\bibfnamefont {F.~N.}\
  \bibnamefont {Baynes}}, \bibinfo {author} {\bibfnamefont {H.~S.}\
  \bibnamefont {Margolis}}, \bibinfo {author} {\bibfnamefont {M.}~\bibnamefont
  {Zampaolo}}, \bibinfo {author} {\bibfnamefont {P.}~\bibnamefont {Thoumany}},
  \bibinfo {author} {\bibfnamefont {M.}~\bibnamefont {Pizzocaro}}, \bibinfo
  {author} {\bibfnamefont {B.}~\bibnamefont {Rauf}}, \bibinfo {author}
  {\bibfnamefont {F.}~\bibnamefont {Bregolin}}, \bibinfo {author}
  {\bibfnamefont {A.}~\bibnamefont {Tampellini}}, \bibinfo {author}
  {\bibfnamefont {P.}~\bibnamefont {Barbieri}}, \bibinfo {author}
  {\bibfnamefont {M.}~\bibnamefont {Zucco}}, \bibinfo {author} {\bibfnamefont
  {G.~A.}\ \bibnamefont {Costanzo}}, \bibinfo {author} {\bibfnamefont
  {C.}~\bibnamefont {Clivati}}, \bibinfo {author} {\bibfnamefont
  {F.}~\bibnamefont {Levi}},\ and\ \bibinfo {author} {\bibfnamefont
  {D.}~\bibnamefont {Calonico}},\ }\bibfield  {title} {\bibinfo {title}
  {Geodesy and metrology with a transportable optical clock},\ }\href
  {https://doi.org/10.1038/s41567-017-0042-3} {\bibfield  {journal} {\bibinfo
  {journal} {Nature Physics}\ }\textbf {\bibinfo {volume} {14}},\ \bibinfo
  {pages} {437} (\bibinfo {year} {2018})}\BibitemShut {NoStop}%
\bibitem [{\citenamefont {Takamoto}\ \emph {et~al.}(2020)\citenamefont
  {Takamoto}, \citenamefont {Ushijima}, \citenamefont {Ohmae}, \citenamefont
  {Yahagi}, \citenamefont {Kokado}, \citenamefont {Shinkai},\ and\
  \citenamefont {Katori}}]{Takamoto2020}%
  \BibitemOpen
  \bibfield  {author} {\bibinfo {author} {\bibfnamefont {M.}~\bibnamefont
  {Takamoto}}, \bibinfo {author} {\bibfnamefont {I.}~\bibnamefont {Ushijima}},
  \bibinfo {author} {\bibfnamefont {N.}~\bibnamefont {Ohmae}}, \bibinfo
  {author} {\bibfnamefont {T.}~\bibnamefont {Yahagi}}, \bibinfo {author}
  {\bibfnamefont {K.}~\bibnamefont {Kokado}}, \bibinfo {author} {\bibfnamefont
  {H.}~\bibnamefont {Shinkai}},\ and\ \bibinfo {author} {\bibfnamefont
  {H.}~\bibnamefont {Katori}},\ }\bibfield  {title} {\bibinfo {title} {Test of
  general relativity by a pair of transportable optical lattice clocks},\
  }\href {https://doi.org/10.1038/s41566-020-0619-8} {\bibfield  {journal}
  {\bibinfo  {journal} {Nature Photonics}\ }\textbf {\bibinfo {volume} {14}},\
  \bibinfo {pages} {411} (\bibinfo {year} {2020})}\BibitemShut {NoStop}%
\bibitem [{\citenamefont {Mehlstäubler}\ \emph {et~al.}(2018)\citenamefont
  {Mehlstäubler}, \citenamefont {Grosche}, \citenamefont {Lisdat},
  \citenamefont {Schmidt},\ and\ \citenamefont {Denker}}]{Mehlstaubler2018}%
  \BibitemOpen
  \bibfield  {author} {\bibinfo {author} {\bibfnamefont {T.~E.}\ \bibnamefont
  {Mehlstäubler}}, \bibinfo {author} {\bibfnamefont {G.}~\bibnamefont
  {Grosche}}, \bibinfo {author} {\bibfnamefont {C.}~\bibnamefont {Lisdat}},
  \bibinfo {author} {\bibfnamefont {P.~O.}\ \bibnamefont {Schmidt}},\ and\
  \bibinfo {author} {\bibfnamefont {H.}~\bibnamefont {Denker}},\ }\bibfield
  {title} {\bibinfo {title} {Atomic clocks for geodesy},\ }\href
  {https://doi.org/10.1088/1361-6633/aab409} {\bibfield  {journal} {\bibinfo
  {journal} {Reports on Progress in Physics}\ }\textbf {\bibinfo {volume}
  {81}},\ \bibinfo {pages} {064401} (\bibinfo {year} {2018})}\BibitemShut
  {NoStop}%
\bibitem [{\citenamefont {Bondarescu}\ \emph {et~al.}(2012)\citenamefont
  {Bondarescu}, \citenamefont {Bondarescu}, \citenamefont {Hetényi},
  \citenamefont {Boschi}, \citenamefont {Jetzer},\ and\ \citenamefont
  {Balakrishna}}]{Bondarescu2012}%
  \BibitemOpen
  \bibfield  {author} {\bibinfo {author} {\bibfnamefont {R.}~\bibnamefont
  {Bondarescu}}, \bibinfo {author} {\bibfnamefont {M.}~\bibnamefont
  {Bondarescu}}, \bibinfo {author} {\bibfnamefont {G.}~\bibnamefont
  {Hetényi}}, \bibinfo {author} {\bibfnamefont {L.}~\bibnamefont {Boschi}},
  \bibinfo {author} {\bibfnamefont {P.}~\bibnamefont {Jetzer}},\ and\ \bibinfo
  {author} {\bibfnamefont {J.}~\bibnamefont {Balakrishna}},\ }\bibfield
  {title} {\bibinfo {title} {{Geophysical applicability of atomic clocks:
  direct continental geoid mapping}},\ }\href
  {https://doi.org/10.1111/j.1365-246X.2012.05636.x} {\bibfield  {journal}
  {\bibinfo  {journal} {Geophysical Journal International}\ }\textbf {\bibinfo
  {volume} {191}},\ \bibinfo {pages} {78} (\bibinfo {year} {2012})}\BibitemShut
  {NoStop}%
\bibitem [{\citenamefont {Denker}\ \emph {et~al.}(2018)\citenamefont {Denker},
  \citenamefont {Timmen}, \citenamefont {Voigt}, \citenamefont {Weyers},
  \citenamefont {Peik}, \citenamefont {Margolis}, \citenamefont {Delva},
  \citenamefont {Wolf},\ and\ \citenamefont {Petit}}]{Denker2018}%
  \BibitemOpen
  \bibfield  {author} {\bibinfo {author} {\bibfnamefont {H.}~\bibnamefont
  {Denker}}, \bibinfo {author} {\bibfnamefont {L.}~\bibnamefont {Timmen}},
  \bibinfo {author} {\bibfnamefont {C.}~\bibnamefont {Voigt}}, \bibinfo
  {author} {\bibfnamefont {S.}~\bibnamefont {Weyers}}, \bibinfo {author}
  {\bibfnamefont {E.}~\bibnamefont {Peik}}, \bibinfo {author} {\bibfnamefont
  {H.~S.}\ \bibnamefont {Margolis}}, \bibinfo {author} {\bibfnamefont
  {P.}~\bibnamefont {Delva}}, \bibinfo {author} {\bibfnamefont
  {P.}~\bibnamefont {Wolf}},\ and\ \bibinfo {author} {\bibfnamefont
  {G.}~\bibnamefont {Petit}},\ }\bibfield  {title} {\bibinfo {title} {Geodetic
  methods to determine the relativistic redshift at the level of $10^{-18}$ in
  the context of international timescales: a review and practical results},\
  }\href {https://doi.org/10.1007/s00190-017-1075-1} {\bibfield  {journal}
  {\bibinfo  {journal} {Journal of Geodesy}\ }\textbf {\bibinfo {volume}
  {92}},\ \bibinfo {pages} {487} (\bibinfo {year} {2018})}\BibitemShut
  {NoStop}%
\bibitem [{\citenamefont {Kolkowitz}\ \emph {et~al.}(2016)\citenamefont
  {Kolkowitz}, \citenamefont {Pikovski}, \citenamefont {Langellier},
  \citenamefont {Lukin}, \citenamefont {Walsworth},\ and\ \citenamefont
  {Ye}}]{Kolkowitz2016}%
  \BibitemOpen
  \bibfield  {author} {\bibinfo {author} {\bibfnamefont {S.}~\bibnamefont
  {Kolkowitz}}, \bibinfo {author} {\bibfnamefont {I.}~\bibnamefont {Pikovski}},
  \bibinfo {author} {\bibfnamefont {N.}~\bibnamefont {Langellier}}, \bibinfo
  {author} {\bibfnamefont {M.~D.}\ \bibnamefont {Lukin}}, \bibinfo {author}
  {\bibfnamefont {R.~L.}\ \bibnamefont {Walsworth}},\ and\ \bibinfo {author}
  {\bibfnamefont {J.}~\bibnamefont {Ye}},\ }\bibfield  {title} {\bibinfo
  {title} {Gravitational wave detection with optical lattice atomic clocks},\
  }\href {https://doi.org/10.1103/PhysRevD.94.124043} {\bibfield  {journal}
  {\bibinfo  {journal} {Phys. Rev. D}\ }\textbf {\bibinfo {volume} {94}},\
  \bibinfo {pages} {124043} (\bibinfo {year} {2016})}\BibitemShut {NoStop}%
\bibitem [{\citenamefont {Su}\ \emph {et~al.}(2018)\citenamefont {Su},
  \citenamefont {Wang}, \citenamefont {Wang},\ and\ \citenamefont
  {Jetzer}}]{Su2018}%
  \BibitemOpen
  \bibfield  {author} {\bibinfo {author} {\bibfnamefont {J.}~\bibnamefont
  {Su}}, \bibinfo {author} {\bibfnamefont {Q.}~\bibnamefont {Wang}}, \bibinfo
  {author} {\bibfnamefont {Q.}~\bibnamefont {Wang}},\ and\ \bibinfo {author}
  {\bibfnamefont {P.}~\bibnamefont {Jetzer}},\ }\bibfield  {title} {\bibinfo
  {title} {Low-frequency gravitational wave detection via double optical clocks
  in space},\ }\href {https://doi.org/10.1088/1361-6382/aab2eb} {\bibfield
  {journal} {\bibinfo  {journal} {Classical and Quantum Gravity}\ }\textbf
  {\bibinfo {volume} {35}},\ \bibinfo {pages} {085010} (\bibinfo {year}
  {2018})}\BibitemShut {NoStop}%
\bibitem [{\citenamefont {Safronova}(2019)}]{Safronova2019}%
  \BibitemOpen
  \bibfield  {author} {\bibinfo {author} {\bibfnamefont {M.~S.}\ \bibnamefont
  {Safronova}},\ }\bibfield  {title} {\bibinfo {title} {The search for
  variation of fundamental constants with clocks},\ }\href
  {https://doi.org/https://doi.org/10.1002/andp.201800364} {\bibfield
  {journal} {\bibinfo  {journal} {Annalen der Physik}\ }\textbf {\bibinfo
  {volume} {531}},\ \bibinfo {pages} {1800364} (\bibinfo {year}
  {2019})}\BibitemShut {NoStop}%
\bibitem [{\citenamefont {Barontini}\ \emph {et~al.}(2022)\citenamefont
  {Barontini}, \citenamefont {Blackburn}, \citenamefont {Boyer}, \citenamefont
  {Butuc-Mayer}, \citenamefont {Calmet}, \citenamefont {Crespo
  L{\'o}pez-Urrutia}, \citenamefont {Curtis}, \citenamefont {Darqui{\'e}},
  \citenamefont {Dunningham}, \citenamefont {Fitch}, \citenamefont {Forgan},
  \citenamefont {Georgiou}, \citenamefont {Gill}, \citenamefont {Godun},
  \citenamefont {Goldwin}, \citenamefont {Guarrera}, \citenamefont {Harwood},
  \citenamefont {Hill}, \citenamefont {Hendricks}, \citenamefont {Jeong},
  \citenamefont {Johnson}, \citenamefont {Keller}, \citenamefont
  {Kozhiparambil~Sajith}, \citenamefont {Kuipers}, \citenamefont {Margolis},
  \citenamefont {Mayo}, \citenamefont {Newman}, \citenamefont {Parsons},
  \citenamefont {Prokhorov}, \citenamefont {Robertson}, \citenamefont
  {Rodewald}, \citenamefont {Safronova}, \citenamefont {Sauer}, \citenamefont
  {Schioppo}, \citenamefont {Sherrill}, \citenamefont {Stadnik}, \citenamefont
  {Szymaniec}, \citenamefont {Tarbutt}, \citenamefont {Thompson}, \citenamefont
  {Tofful}, \citenamefont {Tunesi}, \citenamefont {Vecchio}, \citenamefont
  {Wang},\ and\ \citenamefont {Worm}}]{Barontini2021}%
  \BibitemOpen
  \bibfield  {author} {\bibinfo {author} {\bibfnamefont {G.}~\bibnamefont
  {Barontini}}, \bibinfo {author} {\bibfnamefont {L.}~\bibnamefont
  {Blackburn}}, \bibinfo {author} {\bibfnamefont {V.}~\bibnamefont {Boyer}},
  \bibinfo {author} {\bibfnamefont {F.}~\bibnamefont {Butuc-Mayer}}, \bibinfo
  {author} {\bibfnamefont {X.}~\bibnamefont {Calmet}}, \bibinfo {author}
  {\bibfnamefont {J.~R.}\ \bibnamefont {Crespo L{\'o}pez-Urrutia}}, \bibinfo
  {author} {\bibfnamefont {E.~A.}\ \bibnamefont {Curtis}}, \bibinfo {author}
  {\bibfnamefont {B.}~\bibnamefont {Darqui{\'e}}}, \bibinfo {author}
  {\bibfnamefont {J.}~\bibnamefont {Dunningham}}, \bibinfo {author}
  {\bibfnamefont {N.~J.}\ \bibnamefont {Fitch}}, \bibinfo {author}
  {\bibfnamefont {E.~M.}\ \bibnamefont {Forgan}}, \bibinfo {author}
  {\bibfnamefont {K.}~\bibnamefont {Georgiou}}, \bibinfo {author}
  {\bibfnamefont {P.}~\bibnamefont {Gill}}, \bibinfo {author} {\bibfnamefont
  {R.~M.}\ \bibnamefont {Godun}}, \bibinfo {author} {\bibfnamefont
  {J.}~\bibnamefont {Goldwin}}, \bibinfo {author} {\bibfnamefont
  {V.}~\bibnamefont {Guarrera}}, \bibinfo {author} {\bibfnamefont {A.~C.}\
  \bibnamefont {Harwood}}, \bibinfo {author} {\bibfnamefont {I.~R.}\
  \bibnamefont {Hill}}, \bibinfo {author} {\bibfnamefont {R.~J.}\ \bibnamefont
  {Hendricks}}, \bibinfo {author} {\bibfnamefont {M.}~\bibnamefont {Jeong}},
  \bibinfo {author} {\bibfnamefont {M.~Y.~H.}\ \bibnamefont {Johnson}},
  \bibinfo {author} {\bibfnamefont {M.}~\bibnamefont {Keller}}, \bibinfo
  {author} {\bibfnamefont {L.~P.}\ \bibnamefont {Kozhiparambil~Sajith}},
  \bibinfo {author} {\bibfnamefont {F.}~\bibnamefont {Kuipers}}, \bibinfo
  {author} {\bibfnamefont {H.~S.}\ \bibnamefont {Margolis}}, \bibinfo {author}
  {\bibfnamefont {C.}~\bibnamefont {Mayo}}, \bibinfo {author} {\bibfnamefont
  {P.}~\bibnamefont {Newman}}, \bibinfo {author} {\bibfnamefont {A.~O.}\
  \bibnamefont {Parsons}}, \bibinfo {author} {\bibfnamefont {L.}~\bibnamefont
  {Prokhorov}}, \bibinfo {author} {\bibfnamefont {B.~I.}\ \bibnamefont
  {Robertson}}, \bibinfo {author} {\bibfnamefont {J.}~\bibnamefont {Rodewald}},
  \bibinfo {author} {\bibfnamefont {M.~S.}\ \bibnamefont {Safronova}}, \bibinfo
  {author} {\bibfnamefont {B.~E.}\ \bibnamefont {Sauer}}, \bibinfo {author}
  {\bibfnamefont {M.}~\bibnamefont {Schioppo}}, \bibinfo {author}
  {\bibfnamefont {N.}~\bibnamefont {Sherrill}}, \bibinfo {author}
  {\bibfnamefont {Y.~V.}\ \bibnamefont {Stadnik}}, \bibinfo {author}
  {\bibfnamefont {K.}~\bibnamefont {Szymaniec}}, \bibinfo {author}
  {\bibfnamefont {M.~R.}\ \bibnamefont {Tarbutt}}, \bibinfo {author}
  {\bibfnamefont {R.~C.}\ \bibnamefont {Thompson}}, \bibinfo {author}
  {\bibfnamefont {A.}~\bibnamefont {Tofful}}, \bibinfo {author} {\bibfnamefont
  {J.}~\bibnamefont {Tunesi}}, \bibinfo {author} {\bibfnamefont
  {A.}~\bibnamefont {Vecchio}}, \bibinfo {author} {\bibfnamefont
  {Y.}~\bibnamefont {Wang}},\ and\ \bibinfo {author} {\bibfnamefont
  {S.}~\bibnamefont {Worm}},\ }\bibfield  {title} {\bibinfo {title} {Measuring
  the stability of fundamental constants with a network of clocks},\ }\href
  {https://doi.org/10.1140/epjqt/s40507-022-00130-5} {\bibfield  {journal}
  {\bibinfo  {journal} {EPJ Quantum Technology}\ }\textbf {\bibinfo {volume}
  {9}},\ \bibinfo {pages} {12} (\bibinfo {year} {2022})}\BibitemShut {NoStop}%
\bibitem [{\citenamefont {Lemke}\ \emph {et~al.}(2011)\citenamefont {Lemke},
  \citenamefont {von Stecher}, \citenamefont {Sherman}, \citenamefont {Rey},
  \citenamefont {Oates},\ and\ \citenamefont {Ludlow}}]{Lemke2011}%
  \BibitemOpen
  \bibfield  {author} {\bibinfo {author} {\bibfnamefont {N.~D.}\ \bibnamefont
  {Lemke}}, \bibinfo {author} {\bibfnamefont {J.}~\bibnamefont {von Stecher}},
  \bibinfo {author} {\bibfnamefont {J.~A.}\ \bibnamefont {Sherman}}, \bibinfo
  {author} {\bibfnamefont {A.~M.}\ \bibnamefont {Rey}}, \bibinfo {author}
  {\bibfnamefont {C.~W.}\ \bibnamefont {Oates}},\ and\ \bibinfo {author}
  {\bibfnamefont {A.~D.}\ \bibnamefont {Ludlow}},\ }\bibfield  {title}
  {\bibinfo {title} {$p$-wave cold collisions in an optical lattice clock},\
  }\href {https://doi.org/10.1103/PhysRevLett.107.103902} {\bibfield  {journal}
  {\bibinfo  {journal} {Phys. Rev. Lett.}\ }\textbf {\bibinfo {volume} {107}},\
  \bibinfo {pages} {103902} (\bibinfo {year} {2011})}\BibitemShut {NoStop}%
\bibitem [{\citenamefont {Kobayashi}\ \emph {et~al.}(2020)\citenamefont
  {Kobayashi}, \citenamefont {Akamatsu}, \citenamefont {Hosaka}, \citenamefont
  {Hisai}, \citenamefont {Wada}, \citenamefont {Inaba}, \citenamefont
  {Suzuyama}, \citenamefont {Hong},\ and\ \citenamefont
  {Yasuda}}]{Kobayashi2020}%
  \BibitemOpen
  \bibfield  {author} {\bibinfo {author} {\bibfnamefont {T.}~\bibnamefont
  {Kobayashi}}, \bibinfo {author} {\bibfnamefont {D.}~\bibnamefont {Akamatsu}},
  \bibinfo {author} {\bibfnamefont {K.}~\bibnamefont {Hosaka}}, \bibinfo
  {author} {\bibfnamefont {Y.}~\bibnamefont {Hisai}}, \bibinfo {author}
  {\bibfnamefont {M.}~\bibnamefont {Wada}}, \bibinfo {author} {\bibfnamefont
  {H.}~\bibnamefont {Inaba}}, \bibinfo {author} {\bibfnamefont
  {T.}~\bibnamefont {Suzuyama}}, \bibinfo {author} {\bibfnamefont {F.-L.}\
  \bibnamefont {Hong}},\ and\ \bibinfo {author} {\bibfnamefont
  {M.}~\bibnamefont {Yasuda}},\ }\bibfield  {title} {\bibinfo {title}
  {Demonstration of the nearly continuous operation of an
  {\textsuperscript{171}\uppercase{y}b} optical lattice clock for half a
  year},\ }\href {https://doi.org/10.1088/1681-7575/ab9f1f} {\bibfield
  {journal} {\bibinfo  {journal} {Metrologia}\ }\textbf {\bibinfo {volume}
  {57}},\ \bibinfo {pages} {065021} (\bibinfo {year} {2020})}\BibitemShut
  {NoStop}%
\bibitem [{\citenamefont {Ludlow}\ \emph {et~al.}(2011)\citenamefont {Ludlow},
  \citenamefont {Lemke}, \citenamefont {Sherman}, \citenamefont {Oates},
  \citenamefont {Qu\'em\'ener}, \citenamefont {von Stecher},\ and\
  \citenamefont {Rey}}]{Ludlow2011}%
  \BibitemOpen
  \bibfield  {author} {\bibinfo {author} {\bibfnamefont {A.~D.}\ \bibnamefont
  {Ludlow}}, \bibinfo {author} {\bibfnamefont {N.~D.}\ \bibnamefont {Lemke}},
  \bibinfo {author} {\bibfnamefont {J.~A.}\ \bibnamefont {Sherman}}, \bibinfo
  {author} {\bibfnamefont {C.~W.}\ \bibnamefont {Oates}}, \bibinfo {author}
  {\bibfnamefont {G.}~\bibnamefont {Qu\'em\'ener}}, \bibinfo {author}
  {\bibfnamefont {J.}~\bibnamefont {von Stecher}},\ and\ \bibinfo {author}
  {\bibfnamefont {A.~M.}\ \bibnamefont {Rey}},\ }\bibfield  {title} {\bibinfo
  {title} {Cold-collision-shift cancellation and inelastic scattering in a
  {\uppercase{y}b} optical lattice clock},\ }\href@noop {} {\bibfield
  {journal} {\bibinfo  {journal} {Phys. Rev. A}\ }\textbf {\bibinfo {volume}
  {84}},\ \bibinfo {pages} {052724} (\bibinfo {year} {2011})}\BibitemShut
  {NoStop}%
\bibitem [{\citenamefont {Akatsuka}\ \emph {et~al.}(2010)\citenamefont
  {Akatsuka}, \citenamefont {Takamoto},\ and\ \citenamefont
  {Katori}}]{Akatsuka2010}%
  \BibitemOpen
  \bibfield  {author} {\bibinfo {author} {\bibfnamefont {T.}~\bibnamefont
  {Akatsuka}}, \bibinfo {author} {\bibfnamefont {M.}~\bibnamefont {Takamoto}},\
  and\ \bibinfo {author} {\bibfnamefont {H.}~\bibnamefont {Katori}},\
  }\bibfield  {title} {\bibinfo {title} {Three-dimensional optical lattice
  clock with bosonic $^{88}\mathrm{Sr}$ atoms},\ }\href
  {https://doi.org/10.1103/PhysRevA.81.023402} {\bibfield  {journal} {\bibinfo
  {journal} {Phys. Rev. A}\ }\textbf {\bibinfo {volume} {81}},\ \bibinfo
  {pages} {023402} (\bibinfo {year} {2010})}\BibitemShut {NoStop}%
\bibitem [{\citenamefont {Aeppli}\ \emph {et~al.}(2022)\citenamefont {Aeppli},
  \citenamefont {Chu}, \citenamefont {Bothwell}, \citenamefont {Kennedy},
  \citenamefont {Kedar}, \citenamefont {He}, \citenamefont {Rey},\ and\
  \citenamefont {Ye}}]{Aeppli2022}%
  \BibitemOpen
  \bibfield  {author} {\bibinfo {author} {\bibfnamefont {A.}~\bibnamefont
  {Aeppli}}, \bibinfo {author} {\bibfnamefont {A.}~\bibnamefont {Chu}},
  \bibinfo {author} {\bibfnamefont {T.}~\bibnamefont {Bothwell}}, \bibinfo
  {author} {\bibfnamefont {C.~J.}\ \bibnamefont {Kennedy}}, \bibinfo {author}
  {\bibfnamefont {D.}~\bibnamefont {Kedar}}, \bibinfo {author} {\bibfnamefont
  {P.}~\bibnamefont {He}}, \bibinfo {author} {\bibfnamefont {A.~M.}\
  \bibnamefont {Rey}},\ and\ \bibinfo {author} {\bibfnamefont {J.}~\bibnamefont
  {Ye}},\ }\bibfield  {title} {\bibinfo {title} {Hamiltonian engineering of
  spin-orbit–coupled fermions in a wannier-stark optical lattice clock},\
  }\href@noop {} {\bibfield  {journal} {\bibinfo  {journal} {Science Advances}\
  }\textbf {\bibinfo {volume} {8}},\ \bibinfo {pages} {eadc9242} (\bibinfo
  {year} {2022})}\BibitemShut {NoStop}%
\bibitem [{\citenamefont {Campbell}\ \emph {et~al.}(2017)\citenamefont
  {Campbell}, \citenamefont {Hutson}, \citenamefont {Marti}, \citenamefont
  {Goban}, \citenamefont {Oppong}, \citenamefont {McNally}, \citenamefont
  {Sonderhouse}, \citenamefont {Robinson}, \citenamefont {Zhang}, \citenamefont
  {Bloom},\ and\ \citenamefont {Ye}}]{Campbell2017}%
  \BibitemOpen
  \bibfield  {author} {\bibinfo {author} {\bibfnamefont {S.~L.}\ \bibnamefont
  {Campbell}}, \bibinfo {author} {\bibfnamefont {R.~B.}\ \bibnamefont
  {Hutson}}, \bibinfo {author} {\bibfnamefont {G.~E.}\ \bibnamefont {Marti}},
  \bibinfo {author} {\bibfnamefont {A.}~\bibnamefont {Goban}}, \bibinfo
  {author} {\bibfnamefont {N.~D.}\ \bibnamefont {Oppong}}, \bibinfo {author}
  {\bibfnamefont {R.~L.}\ \bibnamefont {McNally}}, \bibinfo {author}
  {\bibfnamefont {L.}~\bibnamefont {Sonderhouse}}, \bibinfo {author}
  {\bibfnamefont {J.~M.}\ \bibnamefont {Robinson}}, \bibinfo {author}
  {\bibfnamefont {W.}~\bibnamefont {Zhang}}, \bibinfo {author} {\bibfnamefont
  {B.~J.}\ \bibnamefont {Bloom}},\ and\ \bibinfo {author} {\bibfnamefont
  {J.}~\bibnamefont {Ye}},\ }\bibfield  {title} {\bibinfo {title} {A
  fermi-degenerate three-dimensional optical lattice clock},\ }\href
  {https://doi.org/10.1126/science.aam5538} {\bibfield  {journal} {\bibinfo
  {journal} {Science}\ }\textbf {\bibinfo {volume} {358}},\ \bibinfo {pages}
  {90} (\bibinfo {year} {2017})}\BibitemShut {NoStop}%
\bibitem [{\citenamefont {Okaba}\ \emph {et~al.}(2014)\citenamefont {Okaba},
  \citenamefont {Takano}, \citenamefont {Benabid}, \citenamefont {Bradley},
  \citenamefont {Vincetti}, \citenamefont {Maizelis}, \citenamefont
  {Yampol'skii}, \citenamefont {Nori},\ and\ \citenamefont
  {Katori}}]{Okaba2014}%
  \BibitemOpen
  \bibfield  {author} {\bibinfo {author} {\bibfnamefont {S.}~\bibnamefont
  {Okaba}}, \bibinfo {author} {\bibfnamefont {T.}~\bibnamefont {Takano}},
  \bibinfo {author} {\bibfnamefont {F.}~\bibnamefont {Benabid}}, \bibinfo
  {author} {\bibfnamefont {T.}~\bibnamefont {Bradley}}, \bibinfo {author}
  {\bibfnamefont {L.}~\bibnamefont {Vincetti}}, \bibinfo {author}
  {\bibfnamefont {Z.}~\bibnamefont {Maizelis}}, \bibinfo {author}
  {\bibfnamefont {V.}~\bibnamefont {Yampol'skii}}, \bibinfo {author}
  {\bibfnamefont {F.}~\bibnamefont {Nori}},\ and\ \bibinfo {author}
  {\bibfnamefont {H.}~\bibnamefont {Katori}},\ }\bibfield  {title} {\bibinfo
  {title} {Lamb-dicke spectroscopy of atoms in a hollow-core photonic crystal
  fibre},\ }\href {https://doi.org/10.1038/ncomms5096} {\bibfield  {journal}
  {\bibinfo  {journal} {Nature Communications}\ }\textbf {\bibinfo {volume}
  {5}},\ \bibinfo {pages} {4096} (\bibinfo {year} {2014})}\BibitemShut
  {NoStop}%
\bibitem [{\citenamefont {Itano}\ \emph {et~al.}(1993)\citenamefont {Itano},
  \citenamefont {Bergquist}, \citenamefont {Bollinger}, \citenamefont
  {Gilligan}, \citenamefont {Heinzen}, \citenamefont {Moore}, \citenamefont
  {Raizen},\ and\ \citenamefont {Wineland}}]{Itano1993}%
  \BibitemOpen
  \bibfield  {author} {\bibinfo {author} {\bibfnamefont {W.~M.}\ \bibnamefont
  {Itano}}, \bibinfo {author} {\bibfnamefont {J.~C.}\ \bibnamefont
  {Bergquist}}, \bibinfo {author} {\bibfnamefont {J.~J.}\ \bibnamefont
  {Bollinger}}, \bibinfo {author} {\bibfnamefont {J.~M.}\ \bibnamefont
  {Gilligan}}, \bibinfo {author} {\bibfnamefont {D.~J.}\ \bibnamefont
  {Heinzen}}, \bibinfo {author} {\bibfnamefont {F.~L.}\ \bibnamefont {Moore}},
  \bibinfo {author} {\bibfnamefont {M.~G.}\ \bibnamefont {Raizen}},\ and\
  \bibinfo {author} {\bibfnamefont {D.~J.}\ \bibnamefont {Wineland}},\
  }\bibfield  {title} {\bibinfo {title} {Quantum projection noise: Population
  fluctuations in two-level systems},\ }\href
  {https://doi.org/10.1103/PhysRevA.47.3554} {\bibfield  {journal} {\bibinfo
  {journal} {Phys. Rev. A}\ }\textbf {\bibinfo {volume} {47}},\ \bibinfo
  {pages} {3554} (\bibinfo {year} {1993})}\BibitemShut {NoStop}%
\bibitem [{\citenamefont {Bishof}\ \emph {et~al.}(2011)\citenamefont {Bishof},
  \citenamefont {Martin}, \citenamefont {Swallows}, \citenamefont {Benko},
  \citenamefont {Lin}, \citenamefont {Qu\'em\'ener}, \citenamefont {Rey},\ and\
  \citenamefont {Ye}}]{Bishof2011}%
  \BibitemOpen
  \bibfield  {author} {\bibinfo {author} {\bibfnamefont {M.}~\bibnamefont
  {Bishof}}, \bibinfo {author} {\bibfnamefont {M.~J.}\ \bibnamefont {Martin}},
  \bibinfo {author} {\bibfnamefont {M.~D.}\ \bibnamefont {Swallows}}, \bibinfo
  {author} {\bibfnamefont {C.}~\bibnamefont {Benko}}, \bibinfo {author}
  {\bibfnamefont {Y.}~\bibnamefont {Lin}}, \bibinfo {author} {\bibfnamefont
  {G.}~\bibnamefont {Qu\'em\'ener}}, \bibinfo {author} {\bibfnamefont {A.~M.}\
  \bibnamefont {Rey}},\ and\ \bibinfo {author} {\bibfnamefont {J.}~\bibnamefont
  {Ye}},\ }\bibfield  {title} {\bibinfo {title} {Inelastic collisions and
  density-dependent excitation suppression in a ${}^{87}$\uppercase{S}r optical
  lattice clock},\ }\href {https://doi.org/10.1103/PhysRevA.84.052716}
  {\bibfield  {journal} {\bibinfo  {journal} {Phys. Rev. A}\ }\textbf {\bibinfo
  {volume} {84}},\ \bibinfo {pages} {052716} (\bibinfo {year}
  {2011})}\BibitemShut {NoStop}%
\bibitem [{\citenamefont {Alberti}\ \emph {et~al.}(2009)\citenamefont
  {Alberti}, \citenamefont {Ivanov}, \citenamefont {Tino},\ and\ \citenamefont
  {Ferrari}}]{Alberti2009}%
  \BibitemOpen
  \bibfield  {author} {\bibinfo {author} {\bibfnamefont {A.}~\bibnamefont
  {Alberti}}, \bibinfo {author} {\bibfnamefont {V.~V.}\ \bibnamefont {Ivanov}},
  \bibinfo {author} {\bibfnamefont {G.~M.}\ \bibnamefont {Tino}},\ and\
  \bibinfo {author} {\bibfnamefont {G.}~\bibnamefont {Ferrari}},\ }\bibfield
  {title} {\bibinfo {title} {{Engineering the quantum transport of atomic
  wavefunctions over macroscopic distances}},\ }\href
  {https://doi.org/10.1038/nphys1310} {\bibfield  {journal} {\bibinfo
  {journal} {Nature Physics}\ }\textbf {\bibinfo {volume} {5}},\ \bibinfo
  {pages} {547} (\bibinfo {year} {2009})}\BibitemShut {NoStop}%
\bibitem [{\citenamefont {Ivanov}\ \emph {et~al.}(2008)\citenamefont {Ivanov},
  \citenamefont {Alberti}, \citenamefont {Schioppo}, \citenamefont {Ferrari},
  \citenamefont {Artoni}, \citenamefont {Chiofalo},\ and\ \citenamefont
  {Tino}}]{Ivanov2008}%
  \BibitemOpen
  \bibfield  {author} {\bibinfo {author} {\bibfnamefont {V.~V.}\ \bibnamefont
  {Ivanov}}, \bibinfo {author} {\bibfnamefont {A.}~\bibnamefont {Alberti}},
  \bibinfo {author} {\bibfnamefont {M.}~\bibnamefont {Schioppo}}, \bibinfo
  {author} {\bibfnamefont {G.}~\bibnamefont {Ferrari}}, \bibinfo {author}
  {\bibfnamefont {M.}~\bibnamefont {Artoni}}, \bibinfo {author} {\bibfnamefont
  {M.~L.}\ \bibnamefont {Chiofalo}},\ and\ \bibinfo {author} {\bibfnamefont
  {G.~M.}\ \bibnamefont {Tino}},\ }\bibfield  {title} {\bibinfo {title}
  {Coherent delocalization of atomic wave packets in driven lattice
  potentials},\ }\href {https://doi.org/10.1103/PhysRevLett.100.043602}
  {\bibfield  {journal} {\bibinfo  {journal} {Phys. Rev. Lett.}\ }\textbf
  {\bibinfo {volume} {100}},\ \bibinfo {pages} {043602} (\bibinfo {year}
  {2008})}\BibitemShut {NoStop}%
\bibitem [{\citenamefont {Landau}(1932)}]{Landau1932}%
  \BibitemOpen
  \bibfield  {author} {\bibinfo {author} {\bibfnamefont {L.~D.}\ \bibnamefont
  {Landau}},\ }\bibfield  {title} {\bibinfo {title} {Zur theorie der
  energieubertragung \uppercase{II}}\ }(\bibinfo {year} {1932})\BibitemShut
  {NoStop}%
\bibitem [{\citenamefont {Vitanov}\ \emph {et~al.}(2001)\citenamefont
  {Vitanov}, \citenamefont {Halfmann}, \citenamefont {Shore},\ and\
  \citenamefont {Bergmann}}]{Vitanov2001}%
  \BibitemOpen
  \bibfield  {author} {\bibinfo {author} {\bibfnamefont {N.~V.}\ \bibnamefont
  {Vitanov}}, \bibinfo {author} {\bibfnamefont {T.}~\bibnamefont {Halfmann}},
  \bibinfo {author} {\bibfnamefont {B.~W.}\ \bibnamefont {Shore}},\ and\
  \bibinfo {author} {\bibfnamefont {K.}~\bibnamefont {Bergmann}},\ }\bibfield
  {title} {\bibinfo {title} {Laser-induced population transfer by adiabatic
  passage techniques},\ }\href
  {https://doi.org/10.1146/annurev.physchem.52.1.763} {\bibfield  {journal}
  {\bibinfo  {journal} {Annual Review of Physical Chemistry}\ }\textbf
  {\bibinfo {volume} {52}},\ \bibinfo {pages} {763} (\bibinfo {year}
  {2001})}\BibitemShut {NoStop}%
\bibitem [{\citenamefont {Alberti}\ \emph {et~al.}(2010)\citenamefont
  {Alberti}, \citenamefont {Ferrari}, \citenamefont {Ivanov}, \citenamefont
  {Chiofalo},\ and\ \citenamefont {Tino}}]{Alberti2010}%
  \BibitemOpen
  \bibfield  {author} {\bibinfo {author} {\bibfnamefont {A.}~\bibnamefont
  {Alberti}}, \bibinfo {author} {\bibfnamefont {G.}~\bibnamefont {Ferrari}},
  \bibinfo {author} {\bibfnamefont {V.~V.}\ \bibnamefont {Ivanov}}, \bibinfo
  {author} {\bibfnamefont {M.~L.}\ \bibnamefont {Chiofalo}},\ and\ \bibinfo
  {author} {\bibfnamefont {G.~M.}\ \bibnamefont {Tino}},\ }\bibfield  {title}
  {\bibinfo {title} {Atomic wave packets in amplitude-modulated vertical
  optical lattices},\ }\href {https://doi.org/10.1088/1367-2630/12/6/065037}
  {\bibfield  {journal} {\bibinfo  {journal} {New Journal of Physics}\ }\textbf
  {\bibinfo {volume} {12}},\ \bibinfo {pages} {065037} (\bibinfo {year}
  {2010})}\BibitemShut {NoStop}%
\bibitem [{\citenamefont {Chen}(2024)}]{Chen2022}%
  \BibitemOpen
  \bibfield  {author} {\bibinfo {author} {\bibfnamefont {C.}~\bibnamefont
  {Chen}},\ }\bibfield  {title} {\bibinfo {title} {Clock-line-mediated
  \uppercase{S}isyphus cooling},\ }\href@noop {} {\bibfield  {journal}
  {\bibinfo  {journal} {(To Be Published)}\ } (\bibinfo {year}
  {2024})}\BibitemShut {NoStop}%
\bibitem [{\citenamefont {Ushijima}\ \emph {et~al.}(2018)\citenamefont
  {Ushijima}, \citenamefont {Takamoto},\ and\ \citenamefont
  {Katori}}]{Ushijima2018}%
  \BibitemOpen
  \bibfield  {author} {\bibinfo {author} {\bibfnamefont {I.}~\bibnamefont
  {Ushijima}}, \bibinfo {author} {\bibfnamefont {M.}~\bibnamefont {Takamoto}},\
  and\ \bibinfo {author} {\bibfnamefont {H.}~\bibnamefont {Katori}},\
  }\bibfield  {title} {\bibinfo {title} {Operational magic intensity for
  \uppercase{S}r optical lattice clocks},\ }\href
  {https://doi.org/10.1103/PhysRevLett.121.263202} {\bibfield  {journal}
  {\bibinfo  {journal} {Phys. Rev. Lett.}\ }\textbf {\bibinfo {volume} {121}},\
  \bibinfo {pages} {263202} (\bibinfo {year} {2018})}\BibitemShut {NoStop}%
\bibitem [{\citenamefont {Beloy}\ \emph {et~al.}(2020)\citenamefont {Beloy},
  \citenamefont {McGrew}, \citenamefont {Zhang}, \citenamefont {Nicolodi},
  \citenamefont {Fasano}, \citenamefont {Hassan}, \citenamefont {Brown},\ and\
  \citenamefont {Ludlow}}]{Beloy2020}%
  \BibitemOpen
  \bibfield  {author} {\bibinfo {author} {\bibfnamefont {K.}~\bibnamefont
  {Beloy}}, \bibinfo {author} {\bibfnamefont {W.~F.}\ \bibnamefont {McGrew}},
  \bibinfo {author} {\bibfnamefont {X.}~\bibnamefont {Zhang}}, \bibinfo
  {author} {\bibfnamefont {D.}~\bibnamefont {Nicolodi}}, \bibinfo {author}
  {\bibfnamefont {R.~J.}\ \bibnamefont {Fasano}}, \bibinfo {author}
  {\bibfnamefont {Y.~S.}\ \bibnamefont {Hassan}}, \bibinfo {author}
  {\bibfnamefont {R.~C.}\ \bibnamefont {Brown}},\ and\ \bibinfo {author}
  {\bibfnamefont {A.~D.}\ \bibnamefont {Ludlow}},\ }\bibfield  {title}
  {\bibinfo {title} {Modeling motional energy spectra and lattice light shifts
  in optical lattice clocks},\ }\href
  {https://doi.org/10.1103/PhysRevA.101.053416} {\bibfield  {journal} {\bibinfo
   {journal} {Phys. Rev. A}\ }\textbf {\bibinfo {volume} {101}},\ \bibinfo
  {pages} {053416} (\bibinfo {year} {2020})}\BibitemShut {NoStop}%
\bibitem [{\citenamefont {Anderson}\ and\ \citenamefont
  {Kasevich}(1998)}]{Anderson1998}%
  \BibitemOpen
  \bibfield  {author} {\bibinfo {author} {\bibfnamefont {B.~P.}\ \bibnamefont
  {Anderson}}\ and\ \bibinfo {author} {\bibfnamefont {M.~A.}\ \bibnamefont
  {Kasevich}},\ }\bibfield  {title} {\bibinfo {title} {{Macroscopic quantum
  interference from atomic tunnel arrays}},\ }\href
  {https://doi.org/10.1126/science.282.5394.1686} {\bibfield  {journal}
  {\bibinfo  {journal} {Science}\ }\textbf {\bibinfo {volume} {282}},\ \bibinfo
  {pages} {1686} (\bibinfo {year} {1998})}\BibitemShut {NoStop}%
\bibitem [{Sup({\natexlab{a}})}]{Supplemental}%
  \BibitemOpen
  \bibfield  {title} {\bibinfo {title} {See supplemental material at [url will
  be inserted by publisher]},\ }\href@noop {} {\  ({\natexlab{a}})}\BibitemShut
  {NoStop}%
\bibitem [{\citenamefont {Le~Targat}\ \emph {et~al.}(2013)\citenamefont
  {Le~Targat}, \citenamefont {Lorini}, \citenamefont {Le~Coq}, \citenamefont
  {Zawada}, \citenamefont {Gu{\'e}na}, \citenamefont {Abgrall}, \citenamefont
  {Gurov}, \citenamefont {Rosenbusch}, \citenamefont {Rovera}, \citenamefont
  {Nag{\'o}rny}, \citenamefont {Gartman}, \citenamefont {Westergaard},
  \citenamefont {Tobar}, \citenamefont {Lours}, \citenamefont {Santarelli},
  \citenamefont {Clairon}, \citenamefont {Bize}, \citenamefont {Laurent},
  \citenamefont {Lemonde},\ and\ \citenamefont {Lodewyck}}]{LeTargat2013}%
  \BibitemOpen
  \bibfield  {author} {\bibinfo {author} {\bibfnamefont {R.}~\bibnamefont
  {Le~Targat}}, \bibinfo {author} {\bibfnamefont {L.}~\bibnamefont {Lorini}},
  \bibinfo {author} {\bibfnamefont {Y.}~\bibnamefont {Le~Coq}}, \bibinfo
  {author} {\bibfnamefont {M.}~\bibnamefont {Zawada}}, \bibinfo {author}
  {\bibfnamefont {J.}~\bibnamefont {Gu{\'e}na}}, \bibinfo {author}
  {\bibfnamefont {M.}~\bibnamefont {Abgrall}}, \bibinfo {author} {\bibfnamefont
  {M.}~\bibnamefont {Gurov}}, \bibinfo {author} {\bibfnamefont
  {P.}~\bibnamefont {Rosenbusch}}, \bibinfo {author} {\bibfnamefont {D.~G.}\
  \bibnamefont {Rovera}}, \bibinfo {author} {\bibfnamefont {B.}~\bibnamefont
  {Nag{\'o}rny}}, \bibinfo {author} {\bibfnamefont {R.}~\bibnamefont
  {Gartman}}, \bibinfo {author} {\bibfnamefont {P.~G.}\ \bibnamefont
  {Westergaard}}, \bibinfo {author} {\bibfnamefont {M.~E.}\ \bibnamefont
  {Tobar}}, \bibinfo {author} {\bibfnamefont {M.}~\bibnamefont {Lours}},
  \bibinfo {author} {\bibfnamefont {G.}~\bibnamefont {Santarelli}}, \bibinfo
  {author} {\bibfnamefont {A.}~\bibnamefont {Clairon}}, \bibinfo {author}
  {\bibfnamefont {S.}~\bibnamefont {Bize}}, \bibinfo {author} {\bibfnamefont
  {P.}~\bibnamefont {Laurent}}, \bibinfo {author} {\bibfnamefont
  {P.}~\bibnamefont {Lemonde}},\ and\ \bibinfo {author} {\bibfnamefont
  {J.}~\bibnamefont {Lodewyck}},\ }\bibfield  {title} {\bibinfo {title}
  {Experimental realization of an optical second with strontium lattice
  clocks},\ }\href {https://doi.org/10.1038/ncomms3109} {\bibfield  {journal}
  {\bibinfo  {journal} {Nature Communications}\ }\textbf {\bibinfo {volume}
  {4}},\ \bibinfo {pages} {2109} (\bibinfo {year} {2013})}\BibitemShut
  {NoStop}%
\bibitem [{\citenamefont {Bothwell}\ \emph {et~al.}(2022)\citenamefont
  {Bothwell}, \citenamefont {Kennedy}, \citenamefont {Aeppli}, \citenamefont
  {Kedar}, \citenamefont {Robinson}, \citenamefont {Oelker}, \citenamefont
  {Staron},\ and\ \citenamefont {Ye}}]{Bothwell2021}%
  \BibitemOpen
  \bibfield  {author} {\bibinfo {author} {\bibfnamefont {T.}~\bibnamefont
  {Bothwell}}, \bibinfo {author} {\bibfnamefont {C.~J.}\ \bibnamefont
  {Kennedy}}, \bibinfo {author} {\bibfnamefont {A.}~\bibnamefont {Aeppli}},
  \bibinfo {author} {\bibfnamefont {D.}~\bibnamefont {Kedar}}, \bibinfo
  {author} {\bibfnamefont {J.~M.}\ \bibnamefont {Robinson}}, \bibinfo {author}
  {\bibfnamefont {E.}~\bibnamefont {Oelker}}, \bibinfo {author} {\bibfnamefont
  {A.}~\bibnamefont {Staron}},\ and\ \bibinfo {author} {\bibfnamefont
  {J.}~\bibnamefont {Ye}},\ }\bibfield  {title} {\bibinfo {title} {Resolving
  the gravitational redshift across a millimetre-scale atomic sample},\ }\href
  {https://doi.org/10.1038/s41586-021-04349-7} {\bibfield  {journal} {\bibinfo
  {journal} {Nature}\ }\textbf {\bibinfo {volume} {602}},\ \bibinfo {pages}
  {420} (\bibinfo {year} {2022})}\BibitemShut {NoStop}%
\bibitem [{\citenamefont {Xu}\ \emph {et~al.}(2014)\citenamefont {Xu},
  \citenamefont {Singh}, \citenamefont {Zappala}, \citenamefont {Bailey},
  \citenamefont {Dietrich}, \citenamefont {Greene}, \citenamefont {Jiang},
  \citenamefont {Lemke}, \citenamefont {Lu}, \citenamefont {Mueller},\ and\
  \citenamefont {O'Connor}}]{Xu2014}%
  \BibitemOpen
  \bibfield  {author} {\bibinfo {author} {\bibfnamefont {C.-Y.}\ \bibnamefont
  {Xu}}, \bibinfo {author} {\bibfnamefont {J.}~\bibnamefont {Singh}}, \bibinfo
  {author} {\bibfnamefont {J.~C.}\ \bibnamefont {Zappala}}, \bibinfo {author}
  {\bibfnamefont {K.~G.}\ \bibnamefont {Bailey}}, \bibinfo {author}
  {\bibfnamefont {M.~R.}\ \bibnamefont {Dietrich}}, \bibinfo {author}
  {\bibfnamefont {J.~P.}\ \bibnamefont {Greene}}, \bibinfo {author}
  {\bibfnamefont {W.}~\bibnamefont {Jiang}}, \bibinfo {author} {\bibfnamefont
  {N.~D.}\ \bibnamefont {Lemke}}, \bibinfo {author} {\bibfnamefont {Z.-T.}\
  \bibnamefont {Lu}}, \bibinfo {author} {\bibfnamefont {P.}~\bibnamefont
  {Mueller}},\ and\ \bibinfo {author} {\bibfnamefont {T.~P.}\ \bibnamefont
  {O'Connor}},\ }\bibfield  {title} {\bibinfo {title} {Measurement of the
  hyperfine quenching rate of the clock transition in $^{171}\mathrm{Yb}$},\
  }\href {https://doi.org/10.1103/PhysRevLett.113.033003} {\bibfield  {journal}
  {\bibinfo  {journal} {Phys. Rev. Lett.}\ }\textbf {\bibinfo {volume} {113}},\
  \bibinfo {pages} {033003} (\bibinfo {year} {2014})}\BibitemShut {NoStop}%
\bibitem [{Note1()}]{Note1}%
  \BibitemOpen
  \bibinfo {note} {Analytical solutions can be found at \cite
  {Dorscher2018}}\BibitemShut {NoStop}%
\bibitem [{\citenamefont {D\"orscher}\ \emph {et~al.}(2018)\citenamefont
  {D\"orscher}, \citenamefont {Schwarz}, \citenamefont {Al-Masoudi},
  \citenamefont {Falke}, \citenamefont {Sterr},\ and\ \citenamefont
  {Lisdat}}]{Dorscher2018}%
  \BibitemOpen
  \bibfield  {author} {\bibinfo {author} {\bibfnamefont {S.}~\bibnamefont
  {D\"orscher}}, \bibinfo {author} {\bibfnamefont {R.}~\bibnamefont {Schwarz}},
  \bibinfo {author} {\bibfnamefont {A.}~\bibnamefont {Al-Masoudi}}, \bibinfo
  {author} {\bibfnamefont {S.}~\bibnamefont {Falke}}, \bibinfo {author}
  {\bibfnamefont {U.}~\bibnamefont {Sterr}},\ and\ \bibinfo {author}
  {\bibfnamefont {C.}~\bibnamefont {Lisdat}},\ }\bibfield  {title} {\bibinfo
  {title} {Lattice-induced photon scattering in an optical lattice clock},\
  }\href {https://doi.org/10.1103/PhysRevA.97.063419} {\bibfield  {journal}
  {\bibinfo  {journal} {Phys. Rev. A}\ }\textbf {\bibinfo {volume} {97}},\
  \bibinfo {pages} {063419} (\bibinfo {year} {2018})}\BibitemShut {NoStop}%
\bibitem [{\citenamefont {Hutson}\ \emph {et~al.}(2019)\citenamefont {Hutson},
  \citenamefont {Goban}, \citenamefont {Marti}, \citenamefont {Sonderhouse},
  \citenamefont {Sanner},\ and\ \citenamefont {Ye}}]{Hutson2019}%
  \BibitemOpen
  \bibfield  {author} {\bibinfo {author} {\bibfnamefont {R.~B.}\ \bibnamefont
  {Hutson}}, \bibinfo {author} {\bibfnamefont {A.}~\bibnamefont {Goban}},
  \bibinfo {author} {\bibfnamefont {G.~E.}\ \bibnamefont {Marti}}, \bibinfo
  {author} {\bibfnamefont {L.}~\bibnamefont {Sonderhouse}}, \bibinfo {author}
  {\bibfnamefont {C.}~\bibnamefont {Sanner}},\ and\ \bibinfo {author}
  {\bibfnamefont {J.}~\bibnamefont {Ye}},\ }\bibfield  {title} {\bibinfo
  {title} {Engineering quantum states of matter for atomic clocks in shallow
  optical lattices},\ }\href {https://doi.org/10.1103/PhysRevLett.123.123401}
  {\bibfield  {journal} {\bibinfo  {journal} {Phys. Rev. Lett.}\ }\textbf
  {\bibinfo {volume} {123}},\ \bibinfo {pages} {123401} (\bibinfo {year}
  {2019})}\BibitemShut {NoStop}%
\bibitem [{\citenamefont {Norcia}\ \emph {et~al.}(2019)\citenamefont {Norcia},
  \citenamefont {Young}, \citenamefont {Eckner}, \citenamefont {Oelker},
  \citenamefont {Ye},\ and\ \citenamefont {Kaufman}}]{Norcia2019}%
  \BibitemOpen
  \bibfield  {author} {\bibinfo {author} {\bibfnamefont {M.~A.}\ \bibnamefont
  {Norcia}}, \bibinfo {author} {\bibfnamefont {A.~W.}\ \bibnamefont {Young}},
  \bibinfo {author} {\bibfnamefont {W.~J.}\ \bibnamefont {Eckner}}, \bibinfo
  {author} {\bibfnamefont {E.}~\bibnamefont {Oelker}}, \bibinfo {author}
  {\bibfnamefont {J.}~\bibnamefont {Ye}},\ and\ \bibinfo {author}
  {\bibfnamefont {A.~M.}\ \bibnamefont {Kaufman}},\ }\bibfield  {title}
  {\bibinfo {title} {Seconds-scale coherence on an optical clock transition in
  a tweezer array},\ }\href {https://doi.org/10.1126/science.aay0644}
  {\bibfield  {journal} {\bibinfo  {journal} {Science}\ }\textbf {\bibinfo
  {volume} {366}},\ \bibinfo {pages} {93} (\bibinfo {year} {2019})}\BibitemShut
  {NoStop}%
\bibitem [{\citenamefont {Nemitz}\ \emph {et~al.}(2019)\citenamefont {Nemitz},
  \citenamefont {J\o{}rgensen}, \citenamefont {Yanagimoto}, \citenamefont
  {Bregolin},\ and\ \citenamefont {Katori}}]{Nemitz2019}%
  \BibitemOpen
  \bibfield  {author} {\bibinfo {author} {\bibfnamefont {N.}~\bibnamefont
  {Nemitz}}, \bibinfo {author} {\bibfnamefont {A.~A.}\ \bibnamefont
  {J\o{}rgensen}}, \bibinfo {author} {\bibfnamefont {R.}~\bibnamefont
  {Yanagimoto}}, \bibinfo {author} {\bibfnamefont {F.}~\bibnamefont
  {Bregolin}},\ and\ \bibinfo {author} {\bibfnamefont {H.}~\bibnamefont
  {Katori}},\ }\bibfield  {title} {\bibinfo {title} {Modeling light shifts in
  optical lattice clocks},\ }\href {https://doi.org/10.1103/PhysRevA.99.033424}
  {\bibfield  {journal} {\bibinfo  {journal} {Phys. Rev. A}\ }\textbf {\bibinfo
  {volume} {99}},\ \bibinfo {pages} {033424} (\bibinfo {year}
  {2019})}\BibitemShut {NoStop}%
\bibitem [{\citenamefont {Kim}\ \emph {et~al.}(2023)\citenamefont {Kim},
  \citenamefont {Aeppli}, \citenamefont {Bothwell},\ and\ \citenamefont
  {Ye}}]{Kim2022}%
  \BibitemOpen
  \bibfield  {author} {\bibinfo {author} {\bibfnamefont {K.}~\bibnamefont
  {Kim}}, \bibinfo {author} {\bibfnamefont {A.}~\bibnamefont {Aeppli}},
  \bibinfo {author} {\bibfnamefont {T.}~\bibnamefont {Bothwell}},\ and\
  \bibinfo {author} {\bibfnamefont {J.}~\bibnamefont {Ye}},\ }\bibfield
  {title} {\bibinfo {title} {Evaluation of lattice light shift at low
  ${10}^{\ensuremath{-}19}$ uncertainty for a shallow lattice \uppercase{S}r
  optical clock},\ }\href {https://doi.org/10.1103/PhysRevLett.130.113203}
  {\bibfield  {journal} {\bibinfo  {journal} {Phys. Rev. Lett.}\ }\textbf
  {\bibinfo {volume} {130}},\ \bibinfo {pages} {113203} (\bibinfo {year}
  {2023})}\BibitemShut {NoStop}%
\bibitem [{\citenamefont {D\"orscher}\ \emph {et~al.}(2023)\citenamefont
  {D\"orscher}, \citenamefont {Klose}, \citenamefont {Maratha~Palli},\ and\
  \citenamefont {Lisdat}}]{Dorscher2022}%
  \BibitemOpen
  \bibfield  {author} {\bibinfo {author} {\bibfnamefont {S.}~\bibnamefont
  {D\"orscher}}, \bibinfo {author} {\bibfnamefont {J.}~\bibnamefont {Klose}},
  \bibinfo {author} {\bibfnamefont {S.}~\bibnamefont {Maratha~Palli}},\ and\
  \bibinfo {author} {\bibfnamefont {C.}~\bibnamefont {Lisdat}},\ }\bibfield
  {title} {\bibinfo {title} {Experimental determination of the
  $\uppercase{E}2\text{\ensuremath{-}}\uppercase{M}1$ polarizability of the
  strontium clock transition},\ }\href
  {https://doi.org/10.1103/PhysRevResearch.5.L012013} {\bibfield  {journal}
  {\bibinfo  {journal} {Phys. Rev. Res.}\ }\textbf {\bibinfo {volume} {5}},\
  \bibinfo {pages} {L012013} (\bibinfo {year} {2023})}\BibitemShut {NoStop}%
\bibitem [{\citenamefont {Witkowski}\ \emph {et~al.}(2022)\citenamefont
  {Witkowski}, \citenamefont {Bilicki}, \citenamefont {Bober}, \citenamefont
  {Kova\v{c}i\'{c}}, \citenamefont {Singh}, \citenamefont {Tonoyan},\ and\
  \citenamefont {Zawada}}]{Witkowski2022}%
  \BibitemOpen
  \bibfield  {author} {\bibinfo {author} {\bibfnamefont {M.}~\bibnamefont
  {Witkowski}}, \bibinfo {author} {\bibfnamefont {S.}~\bibnamefont {Bilicki}},
  \bibinfo {author} {\bibfnamefont {M.}~\bibnamefont {Bober}}, \bibinfo
  {author} {\bibfnamefont {D.}~\bibnamefont {Kova\v{c}i\'{c}}}, \bibinfo
  {author} {\bibfnamefont {V.}~\bibnamefont {Singh}}, \bibinfo {author}
  {\bibfnamefont {A.}~\bibnamefont {Tonoyan}},\ and\ \bibinfo {author}
  {\bibfnamefont {M.}~\bibnamefont {Zawada}},\ }\bibfield  {title} {\bibinfo
  {title} {Photoionization cross sections of ultracold
  \textsuperscript{88}\uppercase{S}r in
  \textsuperscript{1}\uppercase{P}\textsubscript{1} and
  \textsuperscript{3}\uppercase{S}\textsubscript{1} states at 390 nm and the
  resulting blue-detuned magic wavelength optical lattice clock constraints},\
  }\href {https://doi.org/10.1364/OE.460554} {\bibfield  {journal} {\bibinfo
  {journal} {Opt. Express}\ }\textbf {\bibinfo {volume} {30}},\ \bibinfo
  {pages} {21423} (\bibinfo {year} {2022})}\BibitemShut {NoStop}%
\bibitem [{\citenamefont {Fedorova}\ \emph {et~al.}(2020)\citenamefont
  {Fedorova}, \citenamefont {Golovizin}, \citenamefont {Tregubov},
  \citenamefont {Mishin}, \citenamefont {Provorchenko}, \citenamefont
  {Sorokin}, \citenamefont {Khabarova},\ and\ \citenamefont
  {Kolachevsky}}]{Fedorova2020}%
  \BibitemOpen
  \bibfield  {author} {\bibinfo {author} {\bibfnamefont {E.}~\bibnamefont
  {Fedorova}}, \bibinfo {author} {\bibfnamefont {A.}~\bibnamefont {Golovizin}},
  \bibinfo {author} {\bibfnamefont {D.}~\bibnamefont {Tregubov}}, \bibinfo
  {author} {\bibfnamefont {D.}~\bibnamefont {Mishin}}, \bibinfo {author}
  {\bibfnamefont {D.}~\bibnamefont {Provorchenko}}, \bibinfo {author}
  {\bibfnamefont {V.}~\bibnamefont {Sorokin}}, \bibinfo {author} {\bibfnamefont
  {K.}~\bibnamefont {Khabarova}},\ and\ \bibinfo {author} {\bibfnamefont
  {N.}~\bibnamefont {Kolachevsky}},\ }\bibfield  {title} {\bibinfo {title}
  {Simultaneous preparation of two initial clock states in a thulium optical
  clock},\ }\href {https://doi.org/10.1103/PhysRevA.102.063114} {\bibfield
  {journal} {\bibinfo  {journal} {Phys. Rev. A}\ }\textbf {\bibinfo {volume}
  {102}},\ \bibinfo {pages} {063114} (\bibinfo {year} {2020})}\BibitemShut
  {NoStop}%
\bibitem [{\citenamefont {Muraleedharan}\ \emph {et~al.}(2019)\citenamefont
  {Muraleedharan}, \citenamefont {Miyake},\ and\ \citenamefont
  {Deutsch}}]{Muraleedharan2019}%
  \BibitemOpen
  \bibfield  {author} {\bibinfo {author} {\bibfnamefont {G.}~\bibnamefont
  {Muraleedharan}}, \bibinfo {author} {\bibfnamefont {A.}~\bibnamefont
  {Miyake}},\ and\ \bibinfo {author} {\bibfnamefont {I.~H.}\ \bibnamefont
  {Deutsch}},\ }\bibfield  {title} {\bibinfo {title} {Quantum computational
  supremacy in the sampling of bosonic random walkers on a one-dimensional
  lattice},\ }\href {https://doi.org/10.1088/1367-2630/ab0610} {\bibfield
  {journal} {\bibinfo  {journal} {New Journal of Physics}\ }\textbf {\bibinfo
  {volume} {21}},\ \bibinfo {pages} {055003} (\bibinfo {year}
  {2019})}\BibitemShut {NoStop}%
\bibitem [{\citenamefont {Spar}\ \emph {et~al.}(2022)\citenamefont {Spar},
  \citenamefont {Guardado-Sanchez}, \citenamefont {Chi}, \citenamefont {Yan},\
  and\ \citenamefont {Bakr}}]{Spar2022}%
  \BibitemOpen
  \bibfield  {author} {\bibinfo {author} {\bibfnamefont {B.~M.}\ \bibnamefont
  {Spar}}, \bibinfo {author} {\bibfnamefont {E.}~\bibnamefont
  {Guardado-Sanchez}}, \bibinfo {author} {\bibfnamefont {S.}~\bibnamefont
  {Chi}}, \bibinfo {author} {\bibfnamefont {Z.~Z.}\ \bibnamefont {Yan}},\ and\
  \bibinfo {author} {\bibfnamefont {W.~S.}\ \bibnamefont {Bakr}},\ }\bibfield
  {title} {\bibinfo {title} {Realization of a fermi-hubbard optical tweezer
  array},\ }\href {https://doi.org/10.1103/PhysRevLett.128.223202} {\bibfield
  {journal} {\bibinfo  {journal} {Phys. Rev. Lett.}\ }\textbf {\bibinfo
  {volume} {128}},\ \bibinfo {pages} {223202} (\bibinfo {year}
  {2022})}\BibitemShut {NoStop}%
\bibitem [{\citenamefont {Zheng}\ \emph {et~al.}(2022)\citenamefont {Zheng},
  \citenamefont {Dolde}, \citenamefont {Lochab}, \citenamefont {Merriman},
  \citenamefont {Li},\ and\ \citenamefont {Kolkowitz}}]{Zheng2022}%
  \BibitemOpen
  \bibfield  {author} {\bibinfo {author} {\bibfnamefont {X.}~\bibnamefont
  {Zheng}}, \bibinfo {author} {\bibfnamefont {J.}~\bibnamefont {Dolde}},
  \bibinfo {author} {\bibfnamefont {V.}~\bibnamefont {Lochab}}, \bibinfo
  {author} {\bibfnamefont {B.~N.}\ \bibnamefont {Merriman}}, \bibinfo {author}
  {\bibfnamefont {H.}~\bibnamefont {Li}},\ and\ \bibinfo {author}
  {\bibfnamefont {S.}~\bibnamefont {Kolkowitz}},\ }\bibfield  {title} {\bibinfo
  {title} {Differential clock comparisons with a multiplexed optical lattice
  clock},\ }\href {https://doi.org/10.1038/s41586-021-04344-y} {\bibfield
  {journal} {\bibinfo  {journal} {Nature}\ }\textbf {\bibinfo {volume} {602}},\
  \bibinfo {pages} {425} (\bibinfo {year} {2022})}\BibitemShut {NoStop}%
\bibitem [{\citenamefont {Blatt}\ \emph {et~al.}(2009)\citenamefont {Blatt},
  \citenamefont {Thomsen}, \citenamefont {Campbell}, \citenamefont {Ludlow},
  \citenamefont {Swallows}, \citenamefont {Martin}, \citenamefont {Boyd},\ and\
  \citenamefont {Ye}}]{Blatt2009}%
  \BibitemOpen
  \bibfield  {author} {\bibinfo {author} {\bibfnamefont {S.}~\bibnamefont
  {Blatt}}, \bibinfo {author} {\bibfnamefont {J.~W.}\ \bibnamefont {Thomsen}},
  \bibinfo {author} {\bibfnamefont {G.~K.}\ \bibnamefont {Campbell}}, \bibinfo
  {author} {\bibfnamefont {A.~D.}\ \bibnamefont {Ludlow}}, \bibinfo {author}
  {\bibfnamefont {M.~D.}\ \bibnamefont {Swallows}}, \bibinfo {author}
  {\bibfnamefont {M.~J.}\ \bibnamefont {Martin}}, \bibinfo {author}
  {\bibfnamefont {M.~M.}\ \bibnamefont {Boyd}},\ and\ \bibinfo {author}
  {\bibfnamefont {J.}~\bibnamefont {Ye}},\ }\bibfield  {title} {\bibinfo
  {title} {Rabi spectroscopy and excitation inhomogeneity in a one-dimensional
  optical lattice clock},\ }\href {https://doi.org/10.1103/PhysRevA.80.052703}
  {\bibfield  {journal} {\bibinfo  {journal} {Phys. Rev. A}\ }\textbf {\bibinfo
  {volume} {80}},\ \bibinfo {pages} {052703} (\bibinfo {year}
  {2009})}\BibitemShut {NoStop}%
\bibitem [{\citenamefont {Jalabert}\ and\ \citenamefont
  {Pastawski}(2001)}]{Jalabert2001}%
  \BibitemOpen
  \bibfield  {author} {\bibinfo {author} {\bibfnamefont {R.~A.}\ \bibnamefont
  {Jalabert}}\ and\ \bibinfo {author} {\bibfnamefont {H.~M.}\ \bibnamefont
  {Pastawski}},\ }\bibfield  {title} {\bibinfo {title} {Environment-independent
  decoherence rate in classically chaotic systems},\ }\href
  {https://doi.org/10.1103/PhysRevLett.86.2490} {\bibfield  {journal} {\bibinfo
   {journal} {Phys. Rev. Lett.}\ }\textbf {\bibinfo {volume} {86}},\ \bibinfo
  {pages} {2490} (\bibinfo {year} {2001})}\BibitemShut {NoStop}%
\bibitem [{Sup({\natexlab{b}})}]{SuppCite}%
  \BibitemOpen
  \href@noop {} {\bibinfo {title} {See supplemental material [url] for more
  details on the potential systematic effects of extended samples and raman
  scattering, which includes refs. [55-63].}} ({\natexlab{b}})\BibitemShut
  {NoStop}%
\bibitem [{\citenamefont {Lemonde}\ and\ \citenamefont
  {Wolf}(2005)}]{Lemonde2005}%
  \BibitemOpen
  \bibfield  {author} {\bibinfo {author} {\bibfnamefont {P.}~\bibnamefont
  {Lemonde}}\ and\ \bibinfo {author} {\bibfnamefont {P.}~\bibnamefont {Wolf}},\
  }\bibfield  {title} {\bibinfo {title} {Optical lattice clock with atoms
  confined in a shallow trap},\ }\href
  {https://doi.org/10.1103/PhysRevA.72.033409} {\bibfield  {journal} {\bibinfo
  {journal} {Phys. Rev. A}\ }\textbf {\bibinfo {volume} {72}},\ \bibinfo
  {pages} {033409} (\bibinfo {year} {2005})}\BibitemShut {NoStop}%
\bibitem [{\citenamefont {Blatt}(1967)}]{BLATT1967382}%
  \BibitemOpen
  \bibfield  {author} {\bibinfo {author} {\bibfnamefont {J.~M.}\ \bibnamefont
  {Blatt}},\ }\bibfield  {title} {\bibinfo {title} {Practical points concerning
  the solution of the schrödinger equation},\ }\href
  {https://doi.org/https://doi.org/10.1016/0021-9991(67)90046-0} {\bibfield
  {journal} {\bibinfo  {journal} {Journal of Computational Physics}\ }\textbf
  {\bibinfo {volume} {1}},\ \bibinfo {pages} {382} (\bibinfo {year}
  {1967})}\BibitemShut {NoStop}%
\bibitem [{\citenamefont {Zheng}\ \emph {et~al.}(2023)\citenamefont {Zheng},
  \citenamefont {Dolde}, \citenamefont {Cambria}, \citenamefont {Lim},\ and\
  \citenamefont {Kolkowitz}}]{Zheng2023}%
  \BibitemOpen
  \bibfield  {author} {\bibinfo {author} {\bibfnamefont {X.}~\bibnamefont
  {Zheng}}, \bibinfo {author} {\bibfnamefont {J.}~\bibnamefont {Dolde}},
  \bibinfo {author} {\bibfnamefont {M.~C.}\ \bibnamefont {Cambria}}, \bibinfo
  {author} {\bibfnamefont {H.~M.}\ \bibnamefont {Lim}},\ and\ \bibinfo {author}
  {\bibfnamefont {S.}~\bibnamefont {Kolkowitz}},\ }\bibfield  {title} {\bibinfo
  {title} {A lab-based test of the gravitational redshift with a miniature
  clock network},\ }\href {https://doi.org/10.1038/s41467-023-40629-8}
  {\bibfield  {journal} {\bibinfo  {journal} {Nature Communications}\ }\textbf
  {\bibinfo {volume} {14}},\ \bibinfo {pages} {4886} (\bibinfo {year}
  {2023})}\BibitemShut {NoStop}%
\bibitem [{\citenamefont {Beloy}\ \emph {et~al.}(2014)\citenamefont {Beloy},
  \citenamefont {Hinkley}, \citenamefont {Phillips}, \citenamefont {Sherman},
  \citenamefont {Schioppo}, \citenamefont {Lehman}, \citenamefont {Feldman},
  \citenamefont {Hanssen}, \citenamefont {Oates},\ and\ \citenamefont
  {Ludlow}}]{Beloy2014}%
  \BibitemOpen
  \bibfield  {author} {\bibinfo {author} {\bibfnamefont {K.}~\bibnamefont
  {Beloy}}, \bibinfo {author} {\bibfnamefont {N.}~\bibnamefont {Hinkley}},
  \bibinfo {author} {\bibfnamefont {N.~B.}\ \bibnamefont {Phillips}}, \bibinfo
  {author} {\bibfnamefont {J.~A.}\ \bibnamefont {Sherman}}, \bibinfo {author}
  {\bibfnamefont {M.}~\bibnamefont {Schioppo}}, \bibinfo {author}
  {\bibfnamefont {J.}~\bibnamefont {Lehman}}, \bibinfo {author} {\bibfnamefont
  {A.}~\bibnamefont {Feldman}}, \bibinfo {author} {\bibfnamefont {L.~M.}\
  \bibnamefont {Hanssen}}, \bibinfo {author} {\bibfnamefont {C.~W.}\
  \bibnamefont {Oates}},\ and\ \bibinfo {author} {\bibfnamefont {A.~D.}\
  \bibnamefont {Ludlow}},\ }\bibfield  {title} {\bibinfo {title} {Atomic clock
  with $1\ifmmode\times\else\texttimes\fi{}{10}^{\ensuremath{-}18}$
  room-temperature blackbody stark uncertainty},\ }\href
  {https://doi.org/10.1103/PhysRevLett.113.260801} {\bibfield  {journal}
  {\bibinfo  {journal} {Phys. Rev. Lett.}\ }\textbf {\bibinfo {volume} {113}},\
  \bibinfo {pages} {260801} (\bibinfo {year} {2014})}\BibitemShut {NoStop}%
\bibitem [{\citenamefont {Beloy}\ \emph {et~al.}(2018)\citenamefont {Beloy},
  \citenamefont {Zhang}, \citenamefont {McGrew}, \citenamefont {Hinkley},
  \citenamefont {Yoon}, \citenamefont {Nicolodi}, \citenamefont {Fasano},
  \citenamefont {Sch\"affer}, \citenamefont {Brown},\ and\ \citenamefont
  {Ludlow}}]{Beloy2018}%
  \BibitemOpen
  \bibfield  {author} {\bibinfo {author} {\bibfnamefont {K.}~\bibnamefont
  {Beloy}}, \bibinfo {author} {\bibfnamefont {X.}~\bibnamefont {Zhang}},
  \bibinfo {author} {\bibfnamefont {W.~F.}\ \bibnamefont {McGrew}}, \bibinfo
  {author} {\bibfnamefont {N.}~\bibnamefont {Hinkley}}, \bibinfo {author}
  {\bibfnamefont {T.~H.}\ \bibnamefont {Yoon}}, \bibinfo {author}
  {\bibfnamefont {D.}~\bibnamefont {Nicolodi}}, \bibinfo {author}
  {\bibfnamefont {R.~J.}\ \bibnamefont {Fasano}}, \bibinfo {author}
  {\bibfnamefont {S.~A.}\ \bibnamefont {Sch\"affer}}, \bibinfo {author}
  {\bibfnamefont {R.~C.}\ \bibnamefont {Brown}},\ and\ \bibinfo {author}
  {\bibfnamefont {A.~D.}\ \bibnamefont {Ludlow}},\ }\bibfield  {title}
  {\bibinfo {title} {Faraday-shielded dc stark-shift-free optical lattice
  clock},\ }\href {https://doi.org/10.1103/PhysRevLett.120.183201} {\bibfield
  {journal} {\bibinfo  {journal} {Phys. Rev. Lett.}\ }\textbf {\bibinfo
  {volume} {120}},\ \bibinfo {pages} {183201} (\bibinfo {year}
  {2018})}\BibitemShut {NoStop}%
\bibitem [{\citenamefont {Savukov}\ and\ \citenamefont
  {Johnson}(2002)}]{SavJoh02}%
  \BibitemOpen
  \bibfield  {author} {\bibinfo {author} {\bibfnamefont {I.~M.}\ \bibnamefont
  {Savukov}}\ and\ \bibinfo {author} {\bibfnamefont {W.~R.}\ \bibnamefont
  {Johnson}},\ }\bibfield  {title} {\bibinfo {title} {Combined
  configuration-interaction and many-body-perturbation-theory calculations of
  energy levels and transition amplitudes in be, mg, ca, and sr},\ }\href
  {https://doi.org/10.1103/PhysRevA.65.042503} {\bibfield  {journal} {\bibinfo
  {journal} {Phys. Rev. A}\ }\textbf {\bibinfo {volume} {65}},\ \bibinfo
  {pages} {042503} (\bibinfo {year} {2002})}\BibitemShut {NoStop}%
\bibitem [{\citenamefont {Safronova}\ \emph {et~al.}(2012)\citenamefont
  {Safronova}, \citenamefont {Porsev},\ and\ \citenamefont
  {Clark}}]{SafPorCla12}%
  \BibitemOpen
  \bibfield  {author} {\bibinfo {author} {\bibfnamefont {M.~S.}\ \bibnamefont
  {Safronova}}, \bibinfo {author} {\bibfnamefont {S.~G.}\ \bibnamefont
  {Porsev}},\ and\ \bibinfo {author} {\bibfnamefont {C.~W.}\ \bibnamefont
  {Clark}},\ }\bibfield  {title} {\bibinfo {title} {Ytterbium in quantum gases
  and atomic clocks: van der waals interactions and blackbody shifts},\ }\href
  {https://doi.org/10.1103/PhysRevLett.109.230802} {\bibfield  {journal}
  {\bibinfo  {journal} {Phys. Rev. Lett.}\ }\textbf {\bibinfo {volume} {109}},\
  \bibinfo {pages} {230802} (\bibinfo {year} {2012})}\BibitemShut {NoStop}%
\bibitem [{\citenamefont {Dzuba}\ and\ \citenamefont
  {Derevianko}(2010)}]{DzuDer10}%
  \BibitemOpen
  \bibfield  {author} {\bibinfo {author} {\bibfnamefont {V.~A.}\ \bibnamefont
  {Dzuba}}\ and\ \bibinfo {author} {\bibfnamefont {A.}~\bibnamefont
  {Derevianko}},\ }\bibfield  {title} {\bibinfo {title} {Dynamic
  polarizabilities and related properties of clock states of the ytterbium
  atom},\ }\href {https://doi.org/10.1088/0953-4075/43/7/074011} {\bibfield
  {journal} {\bibinfo  {journal} {J. Phys. B}\ }\textbf {\bibinfo {volume}
  {43}},\ \bibinfo {pages} {074011} (\bibinfo {year} {2010})}\BibitemShut
  {NoStop}%
\bibitem [{\citenamefont {Beloy}(2012)}]{Bel12}%
  \BibitemOpen
  \bibfield  {author} {\bibinfo {author} {\bibfnamefont {K.}~\bibnamefont
  {Beloy}},\ }\bibfield  {title} {\bibinfo {title} {Experimental constraints on
  the polarizabilities of the $6{s}^{2}$
  ${}^{1}\phantom{\rule{-0.16em}{0ex}}{S}_{0}$ and $6s6p$
  ${}^{3}\phantom{\rule{-0.16em}{0ex}}{P}_{0}^{o}$ states of {Yb}},\ }\href
  {https://doi.org/10.1103/PhysRevA.86.022521} {\bibfield  {journal} {\bibinfo
  {journal} {Phys. Rev. A}\ }\textbf {\bibinfo {volume} {86}},\ \bibinfo
  {pages} {022521} (\bibinfo {year} {2012})}\BibitemShut {NoStop}%
\end{thebibliography}
\end{document}